\documentclass[aps,prc,preprint,showpacs,superscriptaddress]{revtex4}
\usepackage{epsfig}
\usepackage{amsmath}
\usepackage[hyperref]{hyperref}

\newcommand{\Pomeron}{I\!\!P}

\begin{document}

\title{Leading twist nuclear shadowing,
nuclear generalized parton distributions,
and nuclear deeply virtual Compton scattering  at small $x$}

\author{K. Goeke}
\email{Klaus.Goeke@tp2.rub.de}
\affiliation{Institut f\"ur Theoretische Physik II, Ruhr-Universit\"at-Bochum,
D-44780 Bochum, Germany}
\author{V. Guzey}
\email{vguzey@jlab.org}
\affiliation{Theory Center, Jefferson Laboratory, Newport News, VA 23606, USA}
\author{M. Siddikov}
\email{marat.siddikov@tp2.rub.de}
\affiliation{Institut f\"ur Theoretische Physik II, Ruhr-Universit\"at-Bochum,
D-44780 Bochum, Germany}
\affiliation{Departamento de F\'isica y Centro de Estudios Subat\'omicos,
Universidad T\'ecnica Federico Santa Mar\'ia, Valpara\'iso, Chile}
\affiliation{Theoretical  Physics Department, Uzbekistan National University, Tashkent 700174, Uzbekistan}

\preprint{USM-TH-243, JLAB-THY-09-941}

\begin{abstract}
We generalize the leading twist theory of nuclear shadowing and 
calculate quark and gluon generalized parton distributions (GPDs) 
of spinless nuclei. 
We predict very large nuclear shadowing for nuclear GPDs.
In the limit of the purely transverse momentum transfer,
our nuclear GPDs become impact-parameter-dependent
nuclear parton distributions (PDFs). 
Nuclear shadowing induces nontrivial correlations between the
impact parameter $b$ and the light-cone fraction $x$.
We make predictions for the deeply virtual Compton scattering (DVCS) 
amplitude and the DVCS cross section on $^{208}$Pb at high energies.
We calculate the cross section of the Bethe-Heitler (BH) process and address the issue of the extraction of the DVCS signal from the 
$e A \to e \gamma A$ cross section.
We find that  the $e A \to e \gamma A$ differential cross section
is dominated by DVCS at the momentum
transfer $t$ near the minima of the nuclear form factor.
We also find that nuclear shadowing leads to dramatic oscillations
of the DVCS beam-spin asymmetry, $A_{LU}$, as a function of $t$.
The position of the points where $A_{LU}$
changes sign is directly related to the magnitude of nuclear shadowing.
\end{abstract}

\maketitle

\section{Introduction}

Hard exclusive reactions and generalized parton distributions (GPDs)
have been in the focus of hadronic physics for the last 
decade~\cite{Mueller:1998fv,Ji:1998pc,Goeke:2001tz,Diehl:2003ny,Belitsky:2005qn,Boffi:2007yc}.
GPDs interpolate between elastic form factors and structure functions and
contain detailed information on distributions and correlations of partons
(quarks and gluons) 
in hadronic targets (pions, nucleons, and nuclei). In particular,
GPDs describe the distribution of partons both in the longitudinal momentum
direction and 
in the impact parameter (transverse) plane~\cite{Burkardt:2002hr} and also allow
to access the total angular momentum of the target carried by the partons~\cite{Ji:1996nm}.

The QCD factorization theorems for hard exclusive 
processes~\cite{Collins:1998be,Collins:1996fb} state that GPDs are
universal distributions that enter the perturbative QCD description of various hard exclusive 
processes such as Deeply Virtual Compton scattering (DVCS), 
$\gamma^{\ast}\, T \to \gamma \,T$
($T$ denotes any hadronic target), exclusive production of mesons, 
$\gamma^{\ast}\, T \to M \,T$ [where $M$ denotes a (pesudo)scalar or a vector meson], and 
many other processes, including generalizations of these two reactions.

Although the factorization theorems make it theoretically possible to extract GPDs from 
the data, this is a difficult task in practice since GPDs are functions of four variables
and the GPDs enter experimentally measured observables in the form of
convolution with hard coefficient functions.
Therefore, there is a clear need for modeling GPDs, both
to interpet the results of the completed experiments in terms of the microscopic structure
of the hadron target and also to plan future experiments.

In this work, 
we study quark and gluon GPDs of heavy nuclei and DVCS on nuclear targets
at small values of Bjorken $x_B$ (large energies). 
In particular, we generalize the theory of leading twist nuclear shadowing~\cite{Frankfurt:1998ym,Frankfurt:2002kd,Frankfurt:2003zd} to the case of GPDs 
and compute next-to-leading order quark and gluon GPDs of nuclei for 
$10^{-5} \leq x_B \leq 0.2$ and at a fixed virtuality $Q^2$.
Using the obtained nuclear 
GPDs, we compute the DVCS amplitude, the DVCS cross section, and 
the DVCS beam-spin asymmetry for the heavy nuclear target of $^{208}$Pb. 
Our results can be summarized as follows:
\begin{itemize}
\item[(i)]
Leading twist nuclear shadowing suppresses very significantly the DVCS 
amplitude and the DVCS cross section at small values of Bjorken $x_B$.
\item[(ii)]
In the $\xi \to 0$ limit, nuclear GPDs reduce to impact-parameter-dependent
nuclear parton distribution functions (PDFs). Therefore, nuclear GPDs allow one
to access the spatial image of nuclear shadowing.
The shadowing correction to nuclear GPDs introduces nontrivial
correlations between the light-cone fraction $x$ and the impact parameter $b$.
\item[(iii)]
DVCS interferes with the purely electromagnetic Bethe-Heitler (BH) process.
At small values of the momentum transfer $t$, which dominate coherent nuclear DVCS
 (without nuclear break-up), and also for the $t$-integrated cross sections,
the BH cross section is much larger than the DVCS one. 
This makes it rather challenging to
extract a small DVCS signal on the background of the
dominant BH contribution to the $e A \to e \gamma A$ cross section.
However, owing to the rapid $t$-dependence,
the DVCS cross section becomes (much) larger than the BH cross section
near the minima of the nuclear form factor.
This suggests that the measurements of nuclear DVCS at the values of $t$ close 
to the minima of the nuclear form factor will not only be very sensitive to the magnitude of
nuclear shadowing (owing to the suppression of the nonshadowed Born contribution), but 
will also have a sufficiently small Bethe-Heitler contribution. 
\item[(iv)]
Another possible way to access nuclear GPDs in the small $x_B$ region is 
through the 
measurement of the DVCS beam-spin asymmetry, $A_{LU}$. 
Nuclear shadowing causes dramatic
oscillations of the asymmetry at the fixed $\phi=90^{\circ}$ as a function of
the momentum transfer $t$. The position of the points where $A_{LU}$
changes sign is directly related to the magnitude of nuclear shadowing.
\end{itemize}

The rest of the paper is structured as follows. In Sec.~\ref{sec:lt_gpds}
we derive the expression for nuclear shadowing for nuclear GPDs.
In Sec.~\ref{sec:image}, we analyze 
the $\xi_A \to 0$ limit of the resulting nuclear GPDs, point out the equivalence
of the nuclear GPDs in this limit to the impact-parameter-dependent 
nuclear PDFs, and discuss
the spacial image of nuclear shadowing.
Predictions for DVCS observables (the DVCS amplitude and cross section and the beam-spin DVCS
asymmetry) are presented and discussed in Sec.~\ref{sec:results}.
Finally, we summarize and draw conclusions in Sec.~\ref{sec:summary}.

\section{Leading twist nuclear shadowing and nuclear GPDs}
\label{sec:lt_gpds}

The nuclear structure function $F_{2A}(x_B,Q^2)$ measured in inclusive 
deep inelastic scattering (DIS) with nuclear targets differs from the sum of free
nucleon structure functions  $F_{2N}(x_B,Q^2)$
over the entire  range of values of Bjorken 
$x_B$~\cite{Frankfurt:1988nt,Arneodo:1992wf,Geesaman:1995yd,Piller:1999wx}.
In particular, for small values of $x_B$, $10^{-5} \leq x_B \leq 0.05-0.1$,
 $F_{2A}(x_B,Q^2)/[A F_{2N}(x_B,Q^2)] <1$, which is called {\it nuclear shadowing}.

As we learned from DIS with fixed nuclear targets,
the effect of nuclear shadowing is quite large for small $x_B$.
The kinematics of the future high-energy collider~\cite{Deshpande:2005wd,eic} 
will cover the small-$x_B$ region, where the effect of nuclear shadowing 
will play a major role.

The leading twist (LT) theory of nuclear shadowing~\cite{Frankfurt:1998ym,Frankfurt:2002kd,Frankfurt:2003zd}
is an approach to nuclear shadowing, in
which nuclear shadowing in DIS with nuclei is explained in terms of hard diffraction
in lepton-nucleon DIS. In particular, by
using the QCD factorization theorems for inclusive and hard diffractive
DIS and generalizing the result for nuclear 
shadowing in pion-deuteron scattering obtained by
V.N.~Gribov~\cite{Gribov:1968jf}, the leading twist theory of nuclear shadowing
makes predictions for the shadowing correction to nuclear PDFs, 
$\delta f_{j/A}(x_B,Q^2) \equiv f_{j/A}(x_B,Q^2)-A f_{j/N}(x_B,Q^2)$,
in terms of the free nucleon (proton) diffractive PDFs
$f_{j/N}^{D(4)}$ for small values of $x_B$, $10^{-5} \leq x_B \leq 0.2$.
One should note that the generalization of Gribov's result
 to DIS and to nuclei heavier than 
deuterium makes an explicit assumption that the diffractive state 
produced in the interaction
with the first nucleon of the target elastically rescatters off the rest of 
the nucleons (quasi-eikonal approximation)~\cite{Frankfurt:1998ym,Frankfurt:2002kd,Frankfurt:2003zd}.
In the limit of low nuclear density, when the interaction with only two nucleons
of the target is important, the relation between
$\delta f_{j/A}(x_B,Q^2)$ and $f_{j/N}^{D(4)}$
is model independent. Since $f_{j/N}^{D(4)}$  
is a leading twist quantity,
so is $\delta f_{j/A}(x_B,Q^2)$, which
explains the name {\it leading twist} theory of nuclear shadowing.

\begin{figure}[t]
\begin{center}
\epsfig{file=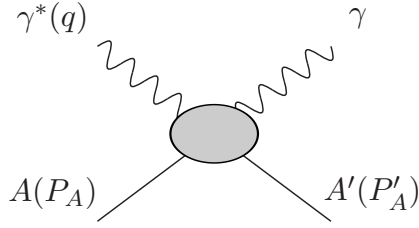,scale=1.1}
\caption{DVCS on a nuclear target.}
\label{fig:DVCS_kinem}
\end{center}
\end{figure}

In this work, we generalize the theory of leading twist nuclear shadowing of usual
nuclear PDFs~\cite{Frankfurt:1998ym,Frankfurt:2002kd,Frankfurt:2003zd}
to the off-forward kinematics, DVCS on nuclear targets, and nuclear GPDs. 
The DVCS amplitude on any hadronic target is defined as a matrix element of the $T$-product of two electromagnetic currents (see, e.g., Ref.~\cite{Goeke:2001tz}),
\begin{equation}
H^{\mu \nu}_A=-i \int d^4 x\, e^{-i\,q \cdot x} \langle P_A^{\prime}| T\{J^{\mu}(x) J^{\nu}(0)\}|P_A \rangle \,, 
\label{eq:ht_1}
\end{equation}
where $q$ ($-q^2=Q^2$) is the momentum of the virtual photon and
$P_A$ and $P_A^{\prime}$ are the momenta of the initial and final nucleus, 
respectively. DVCS on a nuclear target is presented in Fig.~\ref{fig:DVCS_kinem}.

For the analysis of the matrix element in Eq.~(\ref{eq:ht_1}), 
it is convenient
to introduce two light-like vectors $\tilde{p}=1/\sqrt{2}(1,0,0,1)$ 
and $n=1/\sqrt{2}(1,0,0,-1)$ and to work in the so-called symmetric frame, where
$q$ and the average momentum of the initial and 
final nucleus, ${\bar P}_A \equiv (P_A+P_A^{\prime})/2$, are large and 
have no transverse component 
(with respect to the light-like directions defined by $\tilde{p}$ and $n$).
Then, the involved momenta can be parameterized as~\cite{Goeke:2001tz}
\begin{eqnarray}
P_A &=& (1+\xi_A) {\bar P}_A^{+} \,  \tilde{p}+\frac{{\bar M}_A^2}{2\bar{P}_A^{+}}(1-\xi_A)\, n-
\frac{\vec{\Delta}_{\perp}}{2}
 \,, \nonumber\\
P_A^{\prime} &=& (1-\xi_A){\bar P}_A^{+} \, \tilde{p}
+\frac{{\bar M}_A^2}{2\bar{P}_A^{+}}(1+\xi_A)\, n
+\frac{\vec{\Delta}_{\perp}}{2} \,, \nonumber\\
\Delta & \equiv & P_A^{\prime}-P_A= -2 \xi_A {\bar P}_A^{+}\, \tilde{p}+
\xi_A \frac{{\bar M}_A^2}{{\bar P}_A^{+}}\, n+ \
\vec{\Delta}_{\perp} \,, \nonumber\\
q &=& -2 \xi_A {\bar P}_A^{+} \,\tilde{p}+\frac{Q^2}{4\xi_A {\bar P}_A^{+}}\, n \,,
\label{eq:kinem}
\end{eqnarray}
where ${\bar P}_A^{+} \equiv {\bar P}_A \cdot n$;
$\bar{M}_A^2=M_A^2-t/4$, with $M_A$ the mass of the nucleus and
$t=\Delta^2$ the momentum transfer squared;
$Q^2$ is the photon virtuality;
$\vec{\Delta}_{\perp}$ is the component of $\Delta$ orthogonal to the vectors 
$\tilde{p}$ and $n$.
 As follows from the decomposion of Eq.~(\ref{eq:kinem}),
\begin{equation}
\xi_A=\frac{Q^2}{4 \,{\bar P}_A \cdot q}
\approx \frac{x_A}{2-x_A} \,,
\label{eq:xi_A}
\end{equation}
where $x_A$ is the Bjorken variable with respect to the nuclear target,
\begin{equation}
x_A=\frac{Q^2}{2 P_A \cdot q} =\frac{1}{A}\, x_B \,.
\label{eq:xA}
\end{equation}
The Bjorken variable $x_B$ is defined in the usual way with respect to a free nucleon.

In this work, we shall consider spinless nuclei since we are not concerned with
spin effects in nuclear shadowing. 
To the leading twist accuracy and to the leading order in the strong coupling constant,
$H^{\mu \nu}_A$ of a spinless nucleus is expressed in terms of a single
generalized parton distribution, $H_A$, convoluted with the hard scattering
coefficient function $C^{+}(x,\xi_A)$ (see, e.g., Ref.~\cite{Goeke:2001tz}),
\begin{equation}
H^{\mu \nu}_A=-g^{\mu \nu}_{\perp} \int^{1}_{-1} dx \,C^{+}(x,\xi_A)
H_A(x,\xi_A,t,Q^2) \equiv -g^{\mu \nu}_{\perp} \,{\cal H}_A(\xi_A,t,Q^2) \,,
\label{eq:ht_2}
\end{equation}
where $g^{\mu \nu}_{\perp}=g^{\mu \nu}-\tilde{p}^{\mu} n^{\nu}-\tilde{p}^{\nu} n^{\mu}$;
$C^{+}(x,\xi_A)=1/(x-\xi_A+i \epsilon)+1/(x+\xi_A-i \epsilon)$.
The function ${\cal H}_A$ is also called the Compton form factor (CFF). 

\begin{figure}[t]
\begin{center}
\epsfig{file=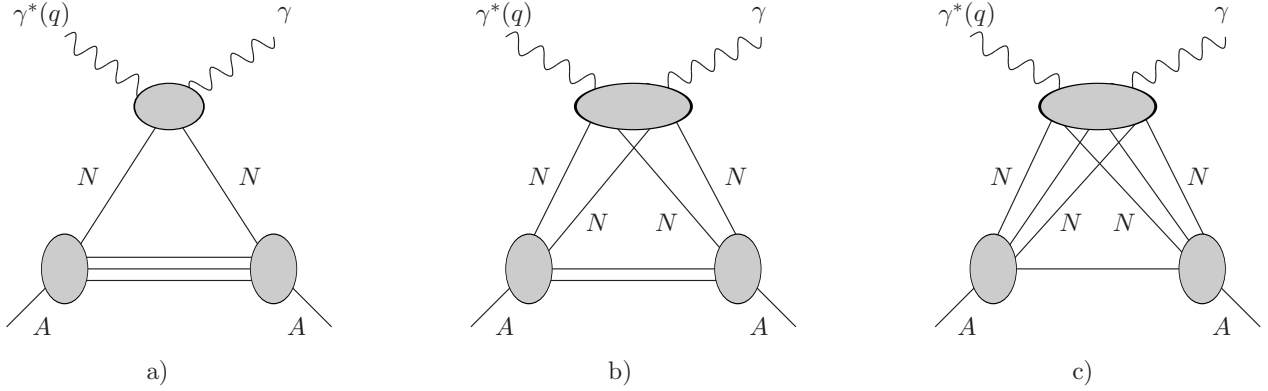,scale=0.875}
\caption{Feynman graphs corresponding to the DVCS amplitude on a nuclear target,
 $H^{\mu \nu}_A$, showing
the impulse (Born) approximation (a) and
the shadowing correction arising from the interaction with
two nucleons (b) and three nucleons of the target (c), respectively.
}
\label{fig:DVCS_NS}
\end{center}
\end{figure}

At sufficiently high energies (small Bjorken $x_B$), 
the virtual photon interacts with many (all) nucleons of the target and 
the DVCS amplitude on a nuclear target,
$H^{\mu \nu}_A$, receives contributions from the graphs presented in 
Fig.~\ref{fig:DVCS_NS}. Figures \ref{fig:DVCS_NS}(a), \ref{fig:DVCS_NS}(b), and \ref{fig:DVCS_NS}(c) correspond to the interaction
with one, two, and three nucleons, respectively.
Graphs that correspond to the interaction with
four and more nucleons of the target are not shown, but they are implied. 
Therefore, $H^{\mu \nu}_A$ can be written as the following sum:
\begin{equation}
H^{\mu \nu}_A=H^{(a)\mu \nu}_A+H^{(b)\mu \nu}_A+H^{(c)\mu \nu}_A+\dots \,,
\label{eq:general}
\end{equation}
where the terms in the right-hand side correspond to the graphs 
shown in Figs.~\ref{fig:DVCS_NS}(a), \ref{fig:DVCS_NS}(b), and \ref{fig:DVCS_NS}(c), respectively.

\subsection{Impulse approximation}

Let us start with the calculation of the graph shown in Fig.~\ref{fig:DVCS_NS}(a).
For the case of a deuterium target, the derivation was done in Ref.~\cite{Cano:2003ju}.
Therefore, in this subsection, we shall follow Ref.~\cite{Cano:2003ju} making straightforward
generalizations to heavier nuclei and high-energy kinematics.
 
The calculation of the graph in Fig.~2(a) can  be carried out straightforwardly using the light-cone (LC) formalism.
In this formalism, each state is characterized by its plus-momentum,
$p^+=p \cdot n=(p^0+p^3)/\sqrt{2}$, the transverse momentum, $\vec{p}_{\perp}$, and
the helicity, $\lambda$.
The minimal Fock component of the nuclear state $|P_A \rangle$ is expressed 
in terms of the nuclear 
LC wave function $\phi_A$ and the product of nucleon states as
\begin{eqnarray}
|P_A^+,\vec{P}_{\perp A} \rangle &=& \sum_{\lambda_i} \int \prod_{i=1}^{A} \frac{d \alpha_i}{\sqrt{\alpha_i}} \frac{d^2 \vec{k}_{\perp i}}{16 \pi^3} 16 \pi^3 
\delta \left(\sum_{j=1}^A \alpha_j-1\right) \delta\left(\sum_{j=1}^A \vec{k}_{\perp j}\right)
\nonumber\\
&\times& \phi_A(\alpha_1,\vec{k}_{\perp 1},\lambda_1,\alpha_2,\vec{k}_{\perp 2},\lambda_2,\dots)|\alpha_i P_A^+,\vec{k}_{\perp i}+\alpha_i\vec{P}_{\perp A},\lambda_i \rangle \,,
\label{eq:lc}
\end{eqnarray}
where $\alpha_i=p_i^+/P_A^+$ is the fraction of the nucleus plus-momentum carried by nucleon $i$.
Since we are not concerned with the correlations of nucleons in
the target nucleus, we take  $\phi_A$ as a product 
of the light-cone wave functions of individual nucleons, $\phi_N$,
\begin{equation}
\phi_A(\alpha_1,\vec{k}_{\perp 1},\lambda_1,\alpha_2,\vec{k}_{\perp 2},\lambda_2,\dots)=\prod_{i=1}^{A} \phi_N(\alpha_i,\vec{k}_{\perp i},\lambda_i) \,.
\label{eq:wf_decomposition}
\end{equation}

Substituting Eq.~(\ref{eq:lc}) for the initial and final nuclear states in
the nuclear DVCS amplitude~[Eq.~(\ref{eq:ht_2})], we obtain
\begin{eqnarray}
H^{(a) \mu \nu}_A &=& - i \int d^4 x\, e^{-i\,q \cdot x} 
\sum_N \sum_{\lambda} \int \frac{d \alpha}{\sqrt{\alpha \,\alpha^{\prime}}}
\frac{d^2 {\vec k}_{\perp}}{16 \pi^3} \,
 \rho_A^N(\alpha^{\prime},\vec{k}_{\perp}^{\prime},\lambda|
\alpha,\vec{k}_{\perp}^{\prime},\lambda) \nonumber\\
& \times& 
\langle p_N^{\prime}| T\{J^{\mu}(x) J^{\nu}(0)\} |p_N \rangle\,,
\label{eq:graph_a}
\end{eqnarray}
where $\sum_N$ denotes the sum over active (interacting) nucleons.
In Eq.~(\ref{eq:graph_a}) and in the rest of the paper, we neglect 
the off-shellness of the nucleons in the photon-nucleon scattering amplitude,
which is a small correction of ${\cal O}(\epsilon/m_N$), where
$\epsilon$ is the average nuclear binding energy and
$m_N$ is the mass of the nucleon.
The effect of the off-shellness in nuclear DVCS was
considered and estimated in Refs.~\cite{Liuti:2005qj,Liuti:2005gi}.

The initial and final states of the active nucleon are
\begin{eqnarray}
|p_N \rangle &=& |\alpha(1+\xi_A) {\bar P}_A^+,
\vec{k}_{\perp}-\alpha \frac{\vec{\Delta}_{\perp}}{2},\lambda  \rangle\,, \nonumber\\
|p_N^{\prime} \rangle &=& |\alpha^{\prime}(1-\xi_A) {\bar P}_A^+,
\vec{k}_{\perp}^{\prime}+\alpha^{\prime} \frac{\vec{\Delta}_{\perp}}{2},\lambda  \rangle \,.
\label{eq:active}
\end{eqnarray}
 The LC fraction and the transverse momentum of the active nucleon are found from the 
conservation of the light-cone energy-momentum in the elementary $\gamma^{\ast}N \to \gamma N$ vertex,
\begin{eqnarray}
\alpha^{\prime} &=& \frac{1+\xi_A}{1-\xi_A}\alpha-\frac{2\xi_A}{1-\xi_A}
\approx \alpha-2\xi_A
 \,,
\nonumber\\
\vec{k}_{\perp}^{\prime} &=& \vec{k}_{\perp}+\frac{1-\alpha}{1-\xi_A} \vec{\Delta}_{\perp} 
\approx \vec{k}_{\perp}+(1-\alpha) \vec{\Delta}_{\perp}
\,.
\label{eq:primed}
\end{eqnarray}
In the above equations, the approximate relations hold after one 
neglects $\xi_A$ compared to unity.

The function $\rho_A^N$ is the overlap between the initial and final nuclear 
LC wave functions,
\begin{align}
\rho_A^N(\alpha^{\prime},&\vec{k}_{\perp}^{\prime},\lambda|
\alpha,\vec{k}_{\perp},\lambda) =\left(\sqrt{\frac{1+\xi_A}{1-\xi_A}}\right)^{A-1}
\phi_N^{\ast}(\alpha^{\prime},\vec{k}_{\perp}^{\prime},\lambda)
\phi_N(\alpha,\vec{k}_{\perp},\lambda)
\nonumber\\
&\times \sum_{\lambda_i}\int \prod_{i=2}^{A} \frac{d \alpha_i \,d^2 \vec{k}_{\perp i}}{16 \pi^3} \delta(\alpha+\sum_{j=2}^A \alpha_j-1) \,16 \pi^3 \delta(\vec{k}_{\perp}+\sum_{j=2}^A \vec{k}_{\perp j})   |\phi_N(\alpha_i,\vec{k}_{\perp i},\lambda_i)|^2 \nonumber\\
&\approx \phi_N^{\ast}(\alpha^{\prime},\vec{k}_{\perp}^{\prime},\lambda)
\phi_N(\alpha,\vec{k}_{\perp},\lambda)
 \,.
\label{eq:rho_tilde}
\end{align}
The last line is an approximation valid for sufficiently large nuclei, when the effects
associated with the motion of the center of mass of the nucleus (taken into account by the 
$\delta$ functions) can be safely neglected.
Note that the helicity conservation requires that the helicity of the active nucleon
be the same in the initial and in the final state.

The matrix element in Eq.~(\ref{eq:graph_a}) can be evaluated by making a transverse boost
to the  symmetric frame of the active nucleon~\cite{Cano:2003ju}. In that frame, 
one can use the standard definition,
\begin{equation}
- i \int d^4 x\, e^{-i\,q \cdot x}
\langle p_N^{\prime}| T\{J^{\mu}(x) J^{\nu}(0)\} |p_N \rangle
 =H_N^{\mu \nu}(\xi_N,t,Q^2)
\,,
\label{eq:graph_a_elem}
\end{equation}
where $H_N^{\mu \nu}$ is the DVCS amplitude for the bound nucleon.
The skewedness $\xi_N$ is determined with the respect to the active nucleon,
\begin{equation}
\xi_N \equiv \frac{Q^2}{4 {\bar p}_N \cdot q}
=\frac{\xi_A}{(1+\xi_A) \alpha-\xi_A} \,,
\label{eq:xi_N}
\end{equation} 
where ${\bar p}_N = (p_N+p_N^{\prime})/2$.
Therefore, we obtain the connection between the DVCS amplitudes for the 
nuclear target and for the bound nucleon,
\begin{equation}
H^{(a) \mu \nu}_A=
\sum_N \sum_{\lambda} \int \frac{d \alpha}{\sqrt{\alpha \,\alpha^{\prime}}}
\frac{d^2 {\vec k}_{\perp}}{16 \pi^3} \,
 \rho_A^N(\alpha^{\prime},\vec{k}_{\perp}^{\prime},\lambda|
\alpha,\vec{k}_{\perp},\lambda) \,H_N^{\mu \nu}(\xi_N,t,Q^2)
\,.
\label{eq:graph_a_2}
\end{equation}

To the leading twist accuracy, the DVCS amplitude for the bound nucleon is 
parametrized in terms of four nucleon GPDs, $H_N$, $E_N$, $\tilde{H}_N$ and
$\tilde{E}_N$:
\begin{align}
H_N^{\mu \nu}(\xi_N,t,Q^2)=&\frac{1}{2 {\bar p}_N^+}(-\tilde{g}^{\mu \nu}_{\perp}) \int^{1}_{-1} dx\, C^+(x,\xi_N)
\Big[H_N(x,\xi_N,t) {\bar u}(p_N^{\prime})\gamma^+ u(p_N) \nonumber\\
&\hspace*{4.5cm}+E_N(x,\xi_N,t) {\bar u}(p_N^{\prime})\frac{i \sigma^{+ \lambda}\Delta_{\lambda}}{2 m_N} u(p_N)\Big] +\dots \,,
\label{eq:ht_nucleon}
\end{align}
where $\dots$ denotes the contribution of the GPDs $\tilde{H}_N$ 
and $\tilde{E}_N$.
The tensor $\tilde{g}^{\mu \nu}_{\perp}$ is defined in the boosted frame 
(the symmetric frame of the active nucleon)
 and, to
a good accuracy, is equal to $g^{\mu \nu}_{\perp}$ entering Eq.~(\ref{eq:ht_2}),
\begin{eqnarray}
\tilde{g}^{\mu \nu}_{\perp} & \equiv & g^{\mu \nu}-\frac{\tilde{q}^{\mu}{\tilde {\bar p}}_N^{\nu}+\tilde{q}^{\nu}{\tilde {\bar p}}_N^{\mu}
}{\tilde{q} \cdot \tilde{p}_N}+\frac{\tilde{q}^{\mu} \tilde{q}^{\nu}}{(\tilde{q} \cdot \tilde{p}_N)^2} \,\tilde{\bar{p}}_N^2+\frac{\tilde{\bar{p}}_N^{\mu} \tilde{q}^{\nu}}{(\tilde{\bar{p}}_N \cdot \tilde{p}_N)^2}\, q^2 \nonumber\\
&\approx& g^{\mu \nu}-\tilde{p}^{\mu} n^{\nu}-\tilde{p}^{\nu} n^{\mu}
+{\cal O}\left(\frac{x_B^2{\bar m}_N^2}{Q^2},\frac{1}{Q^2 R_A^2}\right)
\,,
\label{eq:g_tilde}
\end{eqnarray}
where the vectors $\tilde{q}$ and $\tilde{\bar{p}}_N$ refer to the boosted frame;
$\bar{m}_N^2=m_N^2-t/4$, and $R_A$ is the nuclear radius.
In the derivation of Eq.~(\ref{eq:g_tilde}) we used the fact that the transverse boost 
to the symmetric frame of the active nucleon has not changed the plus-component
of the vectors and that the typical (transverse) momenta of nucleons in a nucleus,
$|\vec{p}_{N \perp}| \sim 1/R_A$, are small compared to the virtuality $Q^2$.

Using the fact that the helicities of the bound
nucleon in the initial and final states are the same and making a natural 
assumption that $\rho_A^N$ is the same for the $\lambda= \pm 1$ helicities,
we observe that the nucleon GPDs $\tilde{H}$ and 
$\tilde{E}$ do not contribute to Eq.~(\ref{eq:graph_a_2}),
which is a consequence of the light-cone spinor algebra (see,
e.g., Ref.~\cite{Cano:2003ju}).
In addition, since we are interested in the kinematics, where the values of $x_B$ and $\xi_N$ are small,
the contribution of the GPDs $E$, which enters  Eq.~(\ref{eq:graph_a_2}) with the prefactor $\xi_N^2$, can be safely neglected. Therefore, we have that
the DVCS amplitude for the bound nucleon reads (keeping in mind the equal helicities of the 
initial and final nucleon)
\begin{equation}
H_N^{\mu \nu}(\xi_N,t,Q^2)=
-g^{\mu \nu}_{\perp}  \,\sqrt{1-\xi_N^2} \,
{\cal H}_N(\xi_N,t,Q^2) 
\approx -g^{\mu \nu}_{\perp}  \, {\cal H}_N(\xi_N,t,Q^2) 
\,,
\label{eq:DVCS_N}
\end{equation}
where ${\cal H}_N$ is the CFF of the bound nucleon.
Thus, we obtain our final relation between the CFF of the nuclear target
in the impulse approximation, ${\cal H}_A^{(a)}$, and that of
the bound nucleon,
\begin{equation}
{\cal H}_A^{(a)}(\xi_A,t,Q^2)=
\sum_N \sum_{\lambda} \int \frac{d \alpha}{\sqrt{\alpha \,\alpha^{\prime}}}
\frac{d^2 {\vec k}_{\perp}}{16 \pi^3} \,
 \rho_A^N(\alpha^{\prime},\vec{k}_{\perp}^{\prime},\lambda|
\alpha,\vec{k}_{\perp},\lambda) \,{\cal H}_N(\xi_N,t,Q^2)
\,.
\label{eq:graph_a_3}
\end{equation}
It is important to point out that the integration over $\alpha$ (longitudinal convolution) and  ${\vec k}_{\perp}$ (transverse convolution)
takes into account the effect of the motion of the bound nucleons in the target 
(Fermi motion effect).
The Fermi motion effect in DVCS on nuclear targets in the form of longitudinal convolution
was also considered in Refs.~\cite{Cano:2003ju,Guzey:2003jh,Scopetta:2004kj,Scopetta:2006wu,Goeke:2008rn}.
Both the longitudinal and transverse convolutions along with the modifications of
the bound nucleon GPDs, which depend on ${\vec k}_{\perp}$, were considered in Refs.~\cite{Liuti:2005qj,Liuti:2005gi}.

To interpret the function $\rho_A^N$ and to fix its normalization, 
it is useful to consider the electromagnetic form factor of a spin-0 
nucleus, $F_A^{\rm e.m.}$, which is defined as the matrix element of the
operator of the electromagnetic current,
\begin{equation}
\langle P_A^{\prime}|J^{\mu}(0)|P_A \rangle=2 {\bar P}_A^{\mu}\, F_A^{\rm e.m.}(t) \,.
\label{eq:ff}
\end{equation}
Using the LC formalism just presented, we consider the plus-component
of Eq.~(\ref{eq:ff}) and obtain
\begin{equation}
2 {\bar P}_A^{+}\,F_A^{\rm e.m.}(t)=
\sum_N \sum_{\lambda} \int \frac{d \alpha}{\sqrt{\alpha \,\alpha^{\prime}}}
\frac{d^2 {\vec k}_{\perp}}{16 \pi^3} \,
 \rho_A^N(\alpha^{\prime},\vec{k}_{\perp}^{\prime},\lambda|
\alpha,\vec{k}_{\perp},\lambda) \langle p_N^{\prime}|J^{+}(0)|p_N \rangle
\,.
\label{eq:ff_2}
\end{equation}
In the reference frame that we work in, the momentum transfer $\Delta$ is predominantly
transverse at small $x_B$ [see Eq.~(\ref{eq:kinem})]. Therefore,
the nucleon matrix element for the same nucleon helicities is (predominantly)
proportional to the Dirac nucleon form factor, $F_{1N}(t)$, 
\begin{equation}
\langle p_N^{\prime}|J^{+}(0)|p_N \rangle \approx {\bar u}(p_N^{\prime}) \gamma^+u(p_N) F_{1N}(t)
 \approx 2 {\bar p}_N^+ F_{1N}(t)=2\,\frac{\xi_A}{\xi_N} \,{\bar P}_A^{+}\,F_{1N}(t) \,.
\label{eq:F1N}
\end{equation}
Therefore,
\begin{equation}
F_A^{\rm e.m.}(t)=
\sum_N F_{1N}(t)
\frac{\xi_A}{\xi_N} \sum_{\lambda} \int 
\frac{d \alpha}{\sqrt{\alpha \, \alpha^{\prime}}}
\frac{d^2 {\vec k}_{\perp}}{16 \pi^3} \,
 \rho_A^N(\alpha^{\prime},\vec{k}_{\perp}^{\prime},\lambda|
\alpha,\vec{k}_{\perp},\lambda) =\sum_N F_A(t) \,F_{1N}(t)
\,,
\label{eq:ff_3}
\end{equation}
where we have introduced the nuclear form factor associated with the distribution
of nucleons in the nucleus (associated with the nuclear density),
\begin{equation}
F_A(t) \equiv \frac{\xi_A}{\xi_N}
\int 
\frac{d \alpha}{\sqrt{\alpha \, \alpha^{\prime}}}
\frac{d^2 {\vec k}_{\perp}}{16 \pi^3} \,
 \rho_A^N(\alpha^{\prime},\vec{k}_{\perp}^{\prime},\lambda|
\alpha,\vec{k}_{\perp},\lambda) \,.
\label{eq:FA}
\end{equation}
As follows from Eq.~(\ref{eq:ff_3}), $F_A(t)$ is normalized to unity 
[$F_A(0)=1$]. This  condition also fixes the normalization of the nuclear LC
wave function,
\begin{equation}
\sum_{\lambda} \int \frac{d \alpha}{\alpha}
\frac{d^2 {\vec k}_{\perp}}{16 \pi^3} \,
 \rho_A^N(\alpha,\vec{k}_{\perp},\lambda|
\alpha,\vec{k}_{\perp},\lambda)=\sum_{\lambda} \int \frac{d \alpha}{\alpha}
\frac{d^2 {\vec k}_{\perp}}{16 \pi^3} \,
 |\phi^N(\alpha,\vec{k}_{\perp},\lambda)|^2=A \,.
\label{eq:lc_normalization}
\end{equation} 

At small $x_B$, the effect of the Fermi motion can be safely neglected
(see, e.g., Ref.~\cite{Smith:2002ci}), and, as a consequence,
Eq.~(\ref{eq:graph_a_3}) can be significantly simplified as follows.
The function $\rho_A^N$ is peaked around $\alpha \approx 1/A$. Thus, if one neglects the 
Fermi motion of the bound nucleon, one evaluates $\xi_N$ at $\alpha=1/A$
(where, for brevity, we shall use the same notation),
\begin{equation}
\xi_N \equiv \xi_N(\alpha=1/A)=\frac{\xi_A}{\frac{1}{A}(1+\xi_A)-\xi_A}
\approx A \,\xi_A \,.
\label{eq:xi_N_approx}
\end{equation}
Therefore, neglecting the Fermi motion and using Eq.~(\ref{eq:FA}), 
Eq.~(\ref{eq:graph_a_3}) can be written in the following simplified and approximate
form:
\begin{equation}
{\cal H}_A^{(a)}(\xi_A,t,Q^2)=\frac{\xi_N}{\xi_A}
\sum_N F_A(t) \,{\cal H}_N(\xi_N,t,Q^2)
\,.
\label{eq:graph_a_4}
\end{equation}
As a number of nucleons, ${\cal H}_A^{(a)}$ scales as $A^2$, which is a natural scaling of the nuclear CFF~\cite{Kirchner:2003wt}.
The inclusion of the Fermi motion effect and the effect associated 
with non-nucleon degrees of the freedom in the nucleus modifies this intuitive 
scaling~\cite{Goeke:2008rn,Polyakov:2002yz}.

\begin{figure}[h]
\begin{center}
\epsfig{file=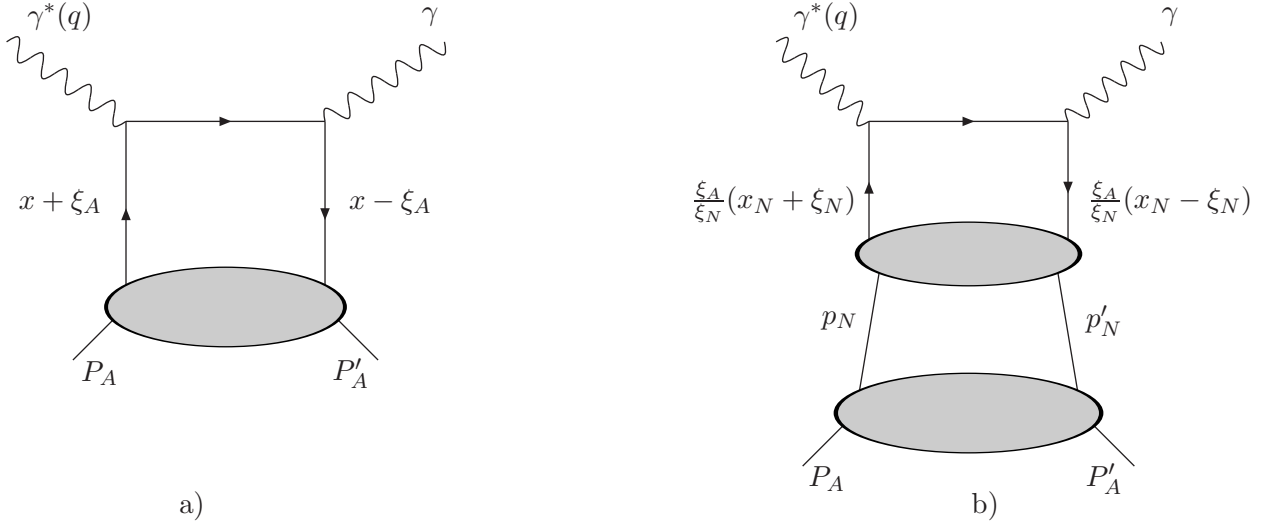,scale=1.0}
\caption{The handbag mechanism for DVCS on a nuclear target.
(a) The generic representation of nuclear GPDs.
(b) A more detailed representation of the same quantity in terms of bound nucleon
GPDs. Shown are relevant quark light-cone fractions and momenta of the active nucleon and the target nucleus.
}
\label{fig:conversion}
\end{center}
\end{figure}

The next important step is the conversion of the relation between nucleus and nucleon 
CFFs [Eq.~(\ref{eq:graph_a_4})] into a similar relation between 
the corresponding GPDs. 
To the leading twist accuracy and to the leading order in the strong coupling constant,
\begin{eqnarray}
{\cal H}_A(\xi_A,t)& =& \int^{1}_{-1} dx\,H_A(x,\xi_A,t) \left(\frac{1}{x-\xi_A+i \epsilon}+
\frac{1}{x+\xi_A-i \epsilon} \right) \,, \nonumber\\
{\cal H}_N(\xi_N,t)& =& \int^{1}_{-1} dx_N\,H_N(x_N,\xi_N,t) \left(\frac{1}{x_N-\xi_N+i \epsilon}+\frac{1}{x_N+\xi_N-i \epsilon} \right) \,.
\label{eq:CFF}
\end{eqnarray}
The relevant quark LC fractions and momenta of the active nucleon and the target nucleus are presented in Fig.~\ref{fig:conversion}. Figure \ref{fig:conversion}(a)
represents the generic handbag approximation for DVCS on a nuclear target,
which expresses the CFF ${\cal H}_A$ in terms of the nuclear GPD $H_A$ and which
corresponds to the first line of Eq.~(\ref{eq:CFF}).

At the same time,  ${\cal H}_A^{(a)}$ can be expressed in terms of the nucleon CFF
${\cal H}_N$ [see Eq.~(\ref{eq:graph_a_4}) and Fig.~\ref{fig:conversion}(b)].
In this case, the nucleon GPD depends on the LC fractions $\xi_N$ defined by 
Eq.~(\ref{eq:xi_N}) and on $x_N$, which is defined with respect to the
active nucleon,
\begin{equation}
x_N \equiv \frac{\bar{k} \cdot n}{{\bar p}_{N} \cdot n}=\frac{x}{\alpha(1+\xi_A)-\xi_A} \,,
\label{eq:x_N}
\end{equation}
where $\bar{k}=(k+k^{\prime})/2$ and $k$ and $k^{\prime}$ are the momenta of the
initial and final lepton, respectively. A useful consequence of Eq.~(\ref{eq:x_N}) is 
the proportionality of the LC fractions $x_N$ and $x$:
\begin{equation}
\frac{x_N}{\xi_N}=\frac{x}{\xi_A} \,.
\label{eq:proportionality}
\end{equation}
This relation allows us to find the LC fractions of the interacting quark 
in Fig.~\ref{fig:conversion}(b), which are equal to
$x+\xi_A=(\xi_A/\xi_N)(x_N+\xi_N)$ and $x-\xi_A=(\xi_A/\xi_N)(x_N-\xi_N)$, respectively.
Since the absolute value of $x_N$ cannot exceed unity, we find
that 
\begin{equation}
|x| \leq \frac{\xi_A}{\xi_N} \approx \frac{1}{A} \,.
\label{eq:x_N_limit}
\end{equation}
Note that the limit $|x| \leq 1/A$ is standard for the approximation, when the nucleus
consists of $A$ stationary nucleons. 
Using Eq.~(\ref{eq:graph_a_3}) and the second line of
Eq.~(\ref{eq:CFF}), we obtain
\begin{align}
{\cal H}_A^{(a)}(\xi_A,t,Q^2)&=\sum_N \sum_{\lambda} \int \frac{d \alpha}{\sqrt{\alpha \,\alpha^{\prime}}}
\frac{d^2 {\vec k}_{\perp}}{16 \pi^3} \,
 \rho_A^N(\alpha^{\prime},\vec{k}_{\perp}^{\prime},\lambda|
\alpha,\vec{k}_{\perp},\lambda) \nonumber\\
&\times
 \int^{1}_{-1} dx_N\,H_N(x_N,\xi_N,t) \left(\frac{1}{x_N-\xi_N+i \epsilon}+\frac{1}{x_N+\xi_N-i \epsilon} \right)
\nonumber\\
&=\sum_N \sum_{\lambda} \int \frac{d \alpha}{\sqrt{\alpha \,\alpha^{\prime}}}
\frac{d^2 {\vec k}_{\perp}}{16 \pi^3} \,
 \rho_A^N(\alpha^{\prime},\vec{k}_{\perp}^{\prime},\lambda|
\alpha,\vec{k}_{\perp},\lambda) \nonumber\\
&\times
 \int^{\xi_A/\xi_N}_{-\xi_A/\xi_N} dx\,H_N(x_N,\xi_N,t) \left(\frac{1}{x-\xi_A+i \epsilon}+\frac{1}{x+\xi_A-i \epsilon} \right)
\,.
\label{eq:graph_a_5}
\end{align}
Recalling the first line of Eq.~(\ref{eq:CFF}) and the limits of integration over 
$x$~[Eq.~(\ref{eq:x_N_limit})], we obtain the desired relation between
the nuclear GPD in the impulse approximation, $H_A^{(a)}$, and the nucleon GPD:
\begin{eqnarray}
H_A^{(a)}(\xi_A,t,Q^2)=
\sum_N \sum_{\lambda} \int \frac{d \alpha}{\sqrt{\alpha \,\alpha^{\prime}}}
\frac{d^2 {\vec k}_{\perp}}{16 \pi^3} \,
 \rho_A^N(\alpha^{\prime},\vec{k}_{\perp}^{\prime},\lambda|
\alpha,\vec{k}_{\perp},\lambda) \,H_N(x_N,\xi_N,t,Q^2)
\,.
\label{eq:graph_a_6}
\end{eqnarray}
We would like to note that Eq.~(\ref{eq:graph_a_6}) could also be derived
starting directly from the definition of the nuclear GPD as the matrix element 
between nuclear states
and applying the LC formalism for the nuclear states, as
we did for the DVCS amplitude above.

Equation~(\ref{eq:graph_a_6}) is derived for the nuclear (nucleon) GPDs, which
are sums of quark GPDs weighted with the quark electric charge squared.
Certainly, the relation between the nuclear and nucleon GPDs holds for individual
parton flavors (quarks and gluons):
\begin{equation}
H_A^{j(a)}(\xi_A,t,Q^2)=
\sum_N \sum_{\lambda} \int \frac{d \alpha}{\sqrt{\alpha \,\alpha^{\prime}}}
\frac{d^2 {\vec k}_{\perp}}{16 \pi^3} \,
 \rho_A^N(\alpha^{\prime},\vec{k}_{\perp}^{\prime},\lambda|
\alpha,\vec{k}_{\perp},\lambda) \,H_N^j(x_N,\xi_N,t,Q^2)
\,,
\label{eq:graph_a_7}
\end{equation}
where $j$ is the parton flavor. 

As we have already explained, the Fermi motion effect can be safely neglected at large energies  [see
Eq.~(\ref{eq:graph_a_4})].
In this case, Eq.~(\ref{eq:graph_a_7}) can be simplified and written in
the following form:
\begin{equation}
H_A^{j(a)}(x,\xi_A,t,Q^2) \approx  \frac{\xi_N}{\xi_A}
\sum_N F_A(t) \,H_N^j(x_N,\xi_N,t,Q^2) \,.
\label{eq:graph_a_7_approx}
\end{equation}

\subsection{Double scattering correction}

\begin{figure}[t]
\begin{center}
\epsfig{file=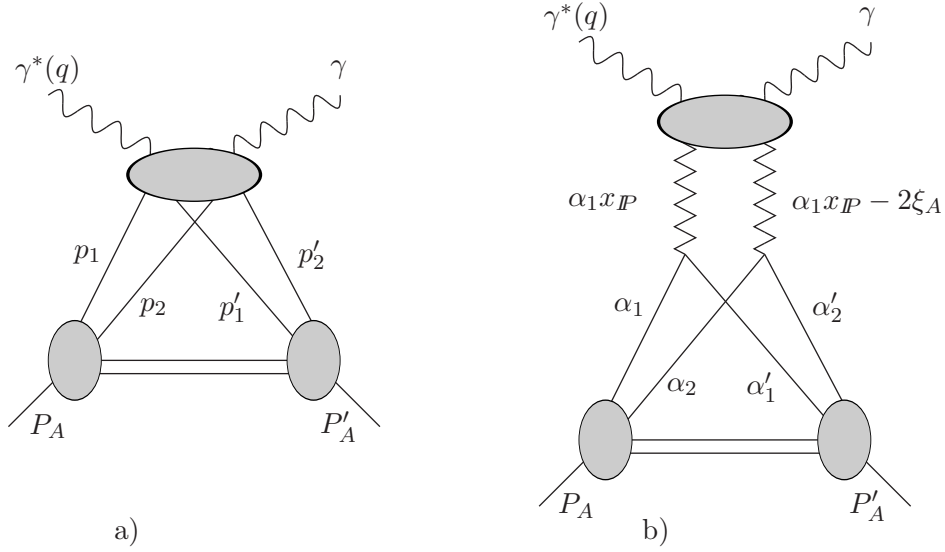,scale=1.0}
\caption{Double rescattering correction to DVCS on a nuclear target.
(a) The shadowing correction in terms of the
$\gamma^{\ast} NN \to \gamma NN$ amplitude. (b) An approximation
based on the assumption that the
shadowing correction can be expressed in terms of DVCS on a Pomeron, 
$\gamma^{\ast} \Pomeron \to \gamma \Pomeron$.
Also shown are the relevant light-cone momentum fractions.
 }
\label{fig:graph_b}
\end{center}
\end{figure}

The graph in Fig.~\ref{fig:DVCS_NS}(b) describes the contribution to DVCS on a nuclear target, when the interaction involves two nucleons of the target. This graph
gives the leading contribution to nuclear shadowing. 
Details of the kinematics of the graph in Fig.~\ref{fig:DVCS_NS}(b) 
 are presented in
Fig.~\ref{fig:graph_b}.

Using the LC formalism, we obtain the following expression
for the contribution of the graph
in Fig.~\ref{fig:DVCS_NS}(b):
\begin{eqnarray}
H_A^{(b) \mu \nu}&=&-i \int d^4 x\,e^{-i q \cdot x} \sum_{{\rm pairs}}
\int \frac{d\alpha_1^{\prime}}{\sqrt{\alpha_1^{\prime} \alpha_2^{\prime}}} \frac{d^2 \vec{k}_{\perp 1}^{\prime}}{16 \pi^3}
 \frac{d\alpha_1 d\alpha_2}{\sqrt{\alpha_1 \alpha_2}} \frac{d^2 \vec{k}_{\perp 1} d^2 \vec{k}_{\perp 2}}{(16 \pi^3)^2}
\nonumber\\
& \times & 
\rho_A^{2N}(\alpha_1^{\prime}\alpha_2^{\prime},\vec{k}_{\perp 1}^{\prime},\vec{k}_{\perp 2}^{\prime}|\alpha_1,\alpha_2,\vec{k}_{\perp 1},\vec{k}_{\perp 2})\,
\langle p_1^{\prime} p_2^{\prime}| T\{J^{\mu}(x) J^{\nu}(0)\} |p_1 p_2 \rangle
\label{eq:graph_b} \,,
\end{eqnarray}
where $\sum_{{\rm pairs}}$ denotes the sum over the pairs of the active nucleons
with  momenta $p_1$ and $p_2$ in the initial state and with  momenta
$p_1^{\prime}$ and $p_2^{\prime}$ in the final state. Each state is characterized
by the corresponding LC fractions and transverse momenta:
\begin{eqnarray}
|p_{1,2} \rangle &=& |\alpha_{1,2}(1+\xi_A) {\bar P}_A^+,
\vec{k}_{\perp 1,2}-\alpha_{1,2} \frac{\vec{\Delta}_{\perp}}{2} \rangle\,, \nonumber\\
|p_{1,2}^{\prime} \rangle &=& |\alpha_{1,2}^{\prime}(1-\xi_A) {\bar P}_A^+,
\vec{k}_{\perp 1,2}^{\prime}+\alpha_{1,2}^{\prime} \frac{\vec{\Delta}_{\perp}}{2} \rangle \,.
\label{eq:active_2}
\end{eqnarray}
The LC fractions and the transverse momenta of the active nucleons are 
related by the conservation of the LC energy-momentum [see also
Eq.~(\ref{eq:primed})]:
\begin{eqnarray}
\alpha_1^{\prime}+\alpha_2^{\prime} &=& \alpha_1+\alpha_2-2\xi_A \,,
\nonumber\\
\vec{k}_{\perp 1}^{\prime}+\vec{k}_{\perp 2}^{\prime}
 &=& \vec{k}_{\perp 1}+\vec{k}_{\perp 2}+\vec{\Delta}_{\perp} \,,
\label{eq:primed_2}
\end{eqnarray}
where we have neglected the factors $\xi_A$ and $\alpha_{1,2}$ compared to unity.

For brevity, we shall not show explicitly the nucleon helicities keeping in mind that
the interaction does not change the helicity of the nucleons.
The function $\rho_A^{2N}$ is given by the following overlap of the nuclear LC
wave functions:
\begin{align}
\rho_A^{2N}(\alpha_1^{\prime}\alpha_2^{\prime},\vec{k}_{\perp 1}^{\prime},\vec{k}_{\perp 2}^{\prime}&|\alpha_1,\alpha_2,\vec{k}_{\perp 1},\vec{k}_{\perp 2})=
\phi_N^{\ast}(\alpha_1^{\prime},\vec{k}_{\perp 1}^{\prime})
\phi_N(\alpha_1,\vec{k}_{\perp 1})
\phi_N^{\ast}(\alpha_2^{\prime},\vec{k}_{\perp 2}^{\prime})
\phi_N(\alpha_2,\vec{k}_{\perp 2}) \nonumber\\
& \times \int \prod_{i=3}^{A} \frac{d \alpha_i \,d^2 \vec{k}_{\perp i}}{16 \pi^3} \delta(\sum_{j=1}^A \alpha_j-1) \,16 \pi^3 \delta(\sum_{j=1}^A \vec{k}_{\perp j})
 \, |\phi_N(\alpha_i^{\prime},\vec{k}_{\perp i}^{\prime})|^2
\nonumber\\
& \approx \phi_N^{\ast}(\alpha_1^{\prime},\vec{k}_{\perp 1}^{\prime})
\phi_N(\alpha_1,\vec{k}_{\perp 1})
\phi_N^{\ast}(\alpha_2^{\prime},\vec{k}_{\perp 2}^{\prime})
\phi_N(\alpha_2,\vec{k}_{\perp 2})
 \,.
\label{eq:rho2_tilde}
\end{align}
Equation~(\ref{eq:graph_b}) is a general expression corresponding to the graph
in Fig.~\ref{fig:DVCS_NS}(b) and to the graph in  Fig.~\ref{fig:graph_b}(a).
To proceed with the derivation, we need to model the multiparticle
matrix element $\langle p_1^{\prime} p_2^{\prime}| T\{J^{\mu}(x) J^{\nu}(0)\} |p_1 p_2 \rangle$. 
Our model for the $\langle p_1^{\prime} p_2^{\prime}| T\{J^{\mu}(x) J^{\nu}(0)\} |p_1 p_2 \rangle$ matrix element is based on the studies of hard inclusive diffraction
in DIS on the proton at HERA in the reaction $e p \to e X p$~\cite{Breitweg:1998gc,Adloff:1997sc,Aktas:2006hy,Aktas:2006hx}, which we shall briefly review in the following.

\begin{figure}[h]
\begin{center}
\epsfig{file=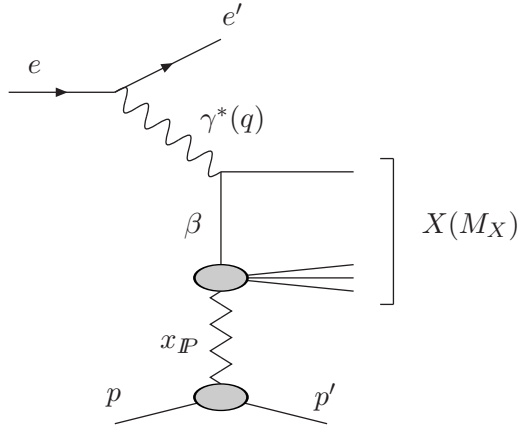,scale=1.}
\caption{Diffractive DIS on the proton.}
\label{fig:diffraction}
\end{center}
\end{figure}

The diffractive DIS $e p \to e X p$ reaction is presented in Fig.~\ref{fig:diffraction}.
The $e p \to e X p$ cross section is expressed in terms of the diffractive 
structure functions $F_2^{D(4)}$ and $F_L^{D(4)}$ as
\begin{equation}
\frac{d^4 \sigma^D_{ep}}{dx_{\Pomeron} dt dx_B dQ^2}=\frac{2 \pi \alpha_{\rm em}^2}{x_B Q^4} \left[(1+(1-y)^2) F_2^{D(4)}(x_B,Q^2,x_{\Pomeron},t) -y^2 F_L^{D(4)}(x_B,Q^2,x_{\Pomeron},t)\right]
\,,
\label{eq:diff_sf}
\end{equation}
where $\alpha_{\rm em}$ is the fine-structure constant and $y=(p \cdot q)/(p\cdot k)$ is the fractional energy loss of the 
incoming lepton. The variables $t$, $x_{\Pomeron}$, and $\beta$ are characteristic for diffractive processes,
\begin{eqnarray}
t&=& (p^{\prime}-p)^2 \,, \nonumber\\
x_{\Pomeron}&=&\frac{q \cdot (p-p^{\prime})}{q \cdot p} \approx \frac{M_X^2+Q^2}{W^2+Q^2} \,, \nonumber\\
\beta &=& \frac{x}{x_{\Pomeron}}=\frac{Q^2}{2q \cdot (p-p^{\prime})} \approx
\frac{Q^2}{Q^2+M_X^2} \,,
\label{eq:diff_kinem}
\end{eqnarray}
where $M_X$ is the invariant mass of the diffractively produced final state and
$W^2=(q+p)^2$. The variable $x_{\Pomeron}$ is the 
fraction of the proton LC momentum lost in the diffractive scattering
(the LC fraction carried by the Pomeron);
 $\beta$ is the LC momentum carried by the interacting
quark (parton).
As follows from the definition of $x_{\Pomeron}$,
the minimal value of $x_{\Pomeron}$ is equal to Bjorken $x_B$,
which corresponds to $M_X=0$.
Typically, the contribution of $F_L^{D(4)}$ is neglected because of its smallness and
 because of
 the kinematic suppression by the $y^2$ factor.

One of the main physics results of HERA is the observation that
hard diffraction in DIS constitutes a large part (10-15\%) of all DIS events
and that hard diffraction in DIS
 is a leading twist phenomenon, that is, that the diffractive
structure function $F_2^{D(4)}$ approximately scales (i.e., it only weakly -- logarithmically --
depends on $Q^2$).

The factorization theorem for hard diffraction in DIS~\cite{Collins:1997sr} states that, at given fixed $t$ and $x_{\Pomeron}$, the diffractive structure function $F_2^{D(4)}$
can be written as convolution of hard scattering coefficient function
$C_j$ with the universal diffractive parton distributions $f_j^{D(4)}$
($j$ is the parton flavor):
\begin{equation}
F_2^{D(4)}(x,Q^2,x_{\Pomeron},t)=\frac{x}{x_{\Pomeron}} \sum_{j=q,\bar{q},g}
\int^1_{x/x_{\Pomeron}} \frac{d \beta^{\prime}}{\beta^{\prime}}
C_j (\frac{x}{x_{\Pomeron} \beta^{\prime}},Q^2) f_j^{D(4)}(\beta^{\prime},Q^2,x_{\Pomeron},t)
\,.
\label{eq:diff_factor}
\end{equation}
It is a phenomenological observation, which follows from
the QCD analysis of the HERA data on inclusive diffraction, 
that the diffractive PDFs $f_j^{D(4)}$ 
can be written as a product 
of the Pomeron flux, $f_{\Pomeron/p}$, 
the parton distribution function of the Pomeron, $f_{j/ \Pomeron}$,
and the factor describing the $t$ dependence,
\begin{equation}
f_j^{D(4)}(\beta,Q^2,x_{\Pomeron},t)=f_{\Pomeron/p}(x_{\Pomeron}) 
f_{j/\Pomeron}(\beta,Q^2) B_{\rm diff} \,e^{B_{\rm diff} t} \,.
\label{eq:diff_Regge_factoriz}
\end{equation}
In Eq.~(\ref{eq:diff_Regge_factoriz}), we neglected the contribution of
the subleading (Reggeon) exchange, which is not important in the considered kinematics.
The Pomeron flux has the following form~\cite{Aktas:2006hy,Aktas:2006hx}
\begin{equation}
f_{\Pomeron/p}(x_{\Pomeron})=\int^{t_{\rm min}}_{-1 \, {\rm GeV}^2} dt \,
A_{\Pomeron}  \frac{e^{B_{\Pomeron}t}}{x_{\Pomeron}^{2\alpha_{\Pomeron}(t)-1}}
\,, \quad \quad \alpha_{\Pomeron}(t)=\alpha_{\Pomeron}(0)+\alpha_{\Pomeron}^{\prime}t \,,
\label{eq:Pomeron_flux}
\end{equation}
where $t_{\rm min} \approx -m_N^2 x_B^2 \approx 0$; 
$B_{\Pomeron}=5.5$ GeV$^{-2}$, $\alpha_{\Pomeron}(0)=1.111$ (Fit B of Ref.~\cite{Aktas:2006hy}), and $\alpha_{\Pomeron}^{\prime}=0.06$ GeV$^{-2}$.
The coefficient $A_{\Pomeron}$ is found from the condition 
$x_{\Pomeron} f_{\Pomeron/p}(x_{\Pomeron})=1$ at $x_{\Pomeron}=0.003$.

The PDFs of the Pomeron, $f_{j/\Pomeron}$, are found from global fits
to the HERA data on hard diffraction taken by the ZEUS and H1 experiments~\cite{Breitweg:1998gc,Adloff:1997sc,Aktas:2006hy,Aktas:2006hx}
 using the QCD factorization 
theorem~[Eq.~(\ref{eq:diff_factor})]. One of the main results of such fits is that
the gluon diffractive PDF is much larger than the quark diffractive PDFs.

The $t$ dependence of hard inclusive diffraction at HERA was recently measured by the
H1 collaboration using the forward proton spectrometer, which allows to detect
the final proton~\cite{Aktas:2006hx}. In the kinematics of the experiment, the 
data was well described by the simple exponential form [Eq.~(\ref{eq:diff_Regge_factoriz})] with the slope $B_{\rm diff} \approx 6$ GeV$^{-2}$.
(Note that $f_j^{D(4)}$ has the dimension GeV$^{-2}$.)

Our model for the $\langle p_1^{\prime} p_2^{\prime}| T\{J^{\mu}(x) J^{\nu}(0)\} |p_1 p_2 \rangle$ matrix element
is based on the observation that,
in the considered kinematics, the interaction of the active nucleons
with the virtual and real photons has a diffractive character and, hence, proceeds
via the $t$-channel exchange with the vacuum quantum numbers (i.e.~the Pomeron).
The model is schematically presented in Fig.~\ref{fig:graph_b}(b).
The space-time picture of the process is the following. Nucleon $1$ with  longitudinal momentum fraction $\alpha_1$ emits a Pomeron with momentum 
fraction $\alpha_1 x_{\Pomeron}$. The virtual photon undergoes DVCS on that Pomeron,
producing a real photon and a Pomeron with the LC  fraction 
$\alpha_1 x_{\Pomeron} -2\xi_A$, which is absorbed by nucleon $2$.
Note that while the skewedness $\xi_A$ is fixed by the external kinematics,
the variable $x_{\Pomeron}$ is integrated over since it is related to 
the LC fractions of 
the active nucleons,
\begin{eqnarray}
\alpha_1^{\prime}&=&\alpha_1-\alpha_1 x_{\Pomeron}\,, \nonumber\\
\alpha_2^{\prime}&=&\alpha_2+\alpha_1 x_{\Pomeron}-2\xi_A \,.
\label{eq:x_Pomeron}
\end{eqnarray}
The variable $x_{\Pomeron}$ has a clear physical interpretation:
it is the fraction of the LC momentum of the nucleon carried by
the Pomeron (see the previous discussion of diffraction in DIS).
Whereas $x_{\Pomeron}$ is the relevant variable for the Pomeron emitted
by nucleon 1, for the Pomeron emitted by nucleon 2, the relevant fraction is
\begin{equation}
\frac{\alpha_2^{\prime}-\alpha_2}{\alpha_2}=\frac{\alpha_1 x_{\Pomeron}-2\xi_A}{\alpha_2}
\approx x_{\Pomeron}-2\xi_N \,.
\label{eq:xP_2}
\end{equation}

Based on this discussion, our model for 
$\langle p_1^{\prime} p_2^{\prime}| T\{J^{\mu}(x) J^{\nu}(0)\} |p_1 p_2 \rangle$
reads
\begin{align}
-i \int d^4 x\,e^{-iq \cdot x} \langle &p_1^{\prime} p_2^{\prime}| T\{J^{\mu}(x) J^{\nu}(0)\} |p_1 p_2 \rangle 
\nonumber\\
&=-(2\pi) k_{\eta}16 \pi B_{\rm diff} \,\phi_{\Pomeron /N}(x_{\Pomeron})
\phi_{\Pomeron /N}(x_{\Pomeron}-2\xi_N) \frac{1}{x_{\Pomeron}} H_{\Pomeron}^{\mu \nu}(\xi_{\Pomeron},t,Q^2)\,,
\label{eq:model_upper}
\end{align}
where $k_{\eta}=(1-i\eta)^2/(1+\eta^2)$,
$\eta \approx \pi/2 (\alpha_{\Pomeron}(0)-1)\approx 0.17$ is the ratio of
the real to imaginary parts of the $\gamma^{\ast} N \to XN$ diffractive amplitude~\cite{Frankfurt:1998ym,Frankfurt:2002kd,Frankfurt:2003zd},
$\phi_{\Pomeron /N}$ is the probability amplitude of emitting a 
Pomeron off the nucleon, and $H_{\Pomeron}^{\mu \nu}$ is the DVCS amplitude
on the Pomeron.
In our analysis, we take $\phi_{\Pomeron /N}(x_{\Pomeron})=\sqrt{f_{\Pomeron /p}(x_{\Pomeron})}$, where
the Pomeron flux is defined by Eq.~(\ref{eq:Pomeron_flux}).
The DVCS amplitude on the Pomeron, $H_{\Pomeron}^{\mu \nu}$, is modeled by
using the PDFs of the Pomeron, $f_{j/\Pomeron}$, which enter 
Eq.~(\ref{eq:diff_Regge_factoriz}). The $t$ dependence of $H_{\Pomeron}^{\mu \nu}$
is given by the factor $e^{B_{\rm diff} t}$.

The skewedness $\xi_{\Pomeron}$ is defined with respect to the Pomeron [compare to Eq.~(\ref{eq:xi_N})],
\begin{equation}
\xi_{\Pomeron}=\frac{Q^2}{4 \,{\bar p}_{\Pomeron} \cdot q}=\frac{\xi_A}{\alpha_1 x_{\Pomeron}-\xi_A} \,,
\label{eq:xi_Pomeron}
\end{equation}
where ${\bar p}_{\Pomeron} =(p_{\Pomeron}+p_{\Pomeron}^{\prime})/2$ with
$p_{\Pomeron}$ and $p_{\Pomeron}^{\prime}$  the momenta of the Pomerons
emitted by nucleon 1 and nucleon 2, respectively.

A few words are in order about the remaining factors in Eq.~(\ref{eq:model_upper}).
The factor of $2 \pi$ comes from the standard definition of the connection between
the Compton scattering amplitude and the structure functions.
The factor of $16\pi$ is specific for diffraction and has its origin in
the optical theorem (see, e.g., Ref.~\cite{Frankfurt:1998ym}).
Note also the overall minus sign, which is a consequence of the fact that 
the considered matrix element is essentially a product of two scattering
amplitudes, which are predominantly imaginary at high-energies.

To implement Eq.~(\ref{eq:model_upper}) in Eq.~(\ref{eq:graph_b}), 
we insert the following identity in Eq.~(\ref{eq:graph_b}):
\begin{equation}
1=\int d \alpha_2^{\prime} \,dx_{\Pomeron}\, \delta(\alpha_2^{\prime}-\alpha_2-\alpha_2(x_{\Pomeron}-2 \xi_N)) \,
\delta(\alpha_1^{\prime}-\alpha_1+\alpha_1x_{\Pomeron})\, \alpha_1 \,.
\label{eq:identity}
\end{equation}
Inserting Eq.~(\ref{eq:model_upper}) in Eq.~(\ref{eq:graph_b}), we 
obtain
\begin{eqnarray}
H_A^{(b) \mu \nu}&=-\Re e &\Big\{\sum_{{\rm pairs}}
\int \frac{d\alpha_1^{\prime}d\alpha_2^{\prime}}{\sqrt{\alpha_1^{\prime} \alpha_2^{\prime}}} \frac{d^2 \vec{k}_{\perp 1}^{\prime}}{16 \pi^3}
 \frac{d\alpha_1 d\alpha_2}{\sqrt{\alpha_1 \alpha_2}} \frac{d^2 \vec{k}_{\perp 1} d^2 \vec{k}_{\perp 2}}{(16 \pi^3)^2} \int_{x_{\Pomeron}^{\min}}^{0.1} dx_{\Pomeron}\, \delta(\alpha_2^{\prime}-\alpha_2-\alpha_2(x_{\Pomeron}-2 \xi_N)) \nonumber\\
& \times &
\delta(\alpha_1^{\prime}-\alpha_1+\alpha_1x_{\Pomeron}) \alpha_1
\phi_N^{\ast}(\alpha_1^{\prime},\vec{k}_{\perp 1}^{\prime})
\phi_N(\alpha_1,\vec{k}_{\perp 1})
\phi_N^{\ast}(\alpha_2^{\prime},\vec{k}_{\perp 2}^{\prime})
\phi_N(\alpha_2,\vec{k}_{\perp 2}) \nonumber\\
&\times& k_{\eta}
(32 \pi^2)B_{\rm diff}\,\phi_{\Pomeron /N}(x_{\Pomeron})
\phi_{\Pomeron /N}(x_{\Pomeron}-2\xi_N)\Big\}\frac{1}{x_{\Pomeron}}
 H_{\Pomeron}^{\mu \nu}(\xi_{\Pomeron},t,Q^2) \,, 
\label{eq:graph_b_2}
\end{eqnarray}
where $x_{\Pomeron}^{\rm min}={\rm max}\{x_N, 2\xi_N\}$.
The limits of integration over $x_{\Pomeron}$ deserve a comment.
The lower limit of integration is the simultaneous requirement that the Pomeron 
LC
fraction in Eq.~(\ref{eq:xP_2}) is non-negative [see also 
Fig.~\ref{fig:graph_b}(b)] and that the Pomeron LC fraction is larger than 
the LC fraction of the active quark, $x_{\Pomeron} \geq x_N$.
The upper limit of integration is the standard condition 
on the produced diffractive masses, which can be cast in the form
$x_{\Pomeron} \leq 0.1$.

In addition, in Eq.~(\ref{eq:graph_b_2}) we made an assumption that multiple 
interactions with the target nucleons lead only to the attenuation of 
$H_A^{(b) \mu \nu}$ and do not introduce an additional imaginary contribution.
This amounts to taking the real part of the expression describing the interaction with
two nucleons of the target.

For the comparison with the predictions of the LT theory of nuclear shadowing
for nuclear PDFs and for the convenience of numerical calculations, we evaluate 
the overlap of the nuclear LC wave functions in Eq.~(\ref{eq:graph_b_2})
in the coordinate space. The Fourier transform of the nuclear LC wave function
reads
\begin{equation}
\phi_N(\alpha,\vec{k}_{\perp })=
\sqrt{2m_N} \int  dz \,d^2 \vec{b}\, e^{im_N\alpha z +i\vec{k}_{\perp} \cdot \vec{b}}\, \phi_N(z,\vec{b}) \,.
\label{eq:wf_coordinate}
\end{equation}
The normalization of the LC wave function in the momentum space [Eq.~(\ref{eq:lc_normalization})] fixes the normalization of the wave function in the coordinate space,
\begin{equation}
\int dz \,d^2 \vec{b}\, |\phi_N(z,\vec{b})|^2 \equiv
\int dz \,d^2 \vec{b}\, \rho_A(z,\vec{b})=
1 \,,
\label{eq:wf_normalization_coordinate}
\end{equation}  
where $\rho_A(z,\vec{b})$ is the nuclear density. 
We have used that $\alpha \approx 1/A$.
In our numerical analysis, we used a two-parameter Fermi form for 
$\rho_A(z,\vec{b})$~\cite{De Jager:1987qc}.

Thus, substituting Eq.~(\ref{eq:wf_coordinate}) into 
Eq.~(\ref{eq:graph_b_2}), using the approximation
\begin{equation}
\frac{\alpha_1}{\sqrt{\alpha^{\prime}_1 \alpha^{\prime}_2 \alpha_1 \alpha_2}} \approx A \approx \frac{\xi_N}{\xi_A} \,,
\end{equation} 
and integrating over the light-cone fractions and the transverse momenta,
we obtain our final expression for $H_A^{(b) \mu \nu}$:
\begin{align}
&H_A^{(b) \mu \nu}=-
\frac{A(A-1)}{2} \frac{\xi_N}{\xi_A}
\, 16 \pi B_{\rm diff} \Re e \Big\{
 \int d^2 \vec{b}\, e^{i \vec{\Delta}_{\perp} \cdot \vec{b}}
\int^{\infty}_{\infty} dz_1 \int^{\infty}_{z_1} dz_2
\int_{x_{\Pomeron}^{\rm min}}^{0.1} dx_{\Pomeron} \,\rho_A(b,z_1) \rho_A(b,z_2)
\nonumber\\
& \times k_{\eta}\, e^{-i m_N z_2 (x_{\Pomeron}-2\xi_N)+i m_N z_1 x_{\Pomeron}}
\phi_{\Pomeron /N}(x_{\Pomeron})
\phi_{\Pomeron /N}(x_{\Pomeron}-2\xi_N) \Big\}\frac{1}{x_{\Pomeron}}
H_{\Pomeron}^{\mu \nu}(\xi_{\Pomeron},t_{\rm min},Q^2) \,, 
\label{eq:graph_b_3}
\end{align}
where we have used that $\sum_{\rm pairs}=A(A-1)/2$.
Note that to perform the Fourier transform, 
we neglected the weak  $t$ dependence of $H_{\Pomeron}^{\mu \nu}$ compared to the 
rapid $t$ dependence of the nuclear distribution and, hence, evaluated
$H_{\Pomeron}^{\mu \nu}$ at the minimal momentum transfer $t_{\rm min} \approx -m_N^2 x_B^2 \approx 0$.
We also introduced the $z_2 > z_1$ ordering
to reflect the space-time evolution of the 
$\gamma^{\ast}NN \to \gamma NN$ scattering 
(see also, e.g., Ref.~\cite{Bauer:1977iq}).

Equation~(\ref{eq:graph_b_3}) can be turned into the relation between the nuclear GPD
and GPD of the Pomeron, quite similarly to the corresponding derivation in the
previous section. The DVCS amplitude on the Pomeron, $H_{\Pomeron}^{\mu \nu}$, 
is expressed in terms of the CFF of the Pomeron,
${\cal H}_{\Pomeron}$, as
\begin{equation}
H_{\Pomeron}^{\mu \nu}(\xi_{\Pomeron},t,Q^2) \approx -g_{\perp}^{\mu \nu} {\cal H}_{\Pomeron}
(\xi_{\Pomeron},t,Q^2) \,,
\label{eq:Pomeron_CFF}
\end{equation}
where we neglected the same terms as in Eq.~(\ref{eq:g_tilde}).
Therefore, for the contribution of the graph in Fig.~\ref{fig:DVCS_NS}(b) to the nuclear 
CFF we obtain
\begin{align}
&{\cal H}_A^{(b)}=-
\frac{A(A-1)}{2} \frac{\xi_N}{\xi_A}
\, 16 \pi B_{\rm diff} \Re e \Big\{
 \int d^2 \vec{b}\, e^{i \vec{\Delta}_{\perp} \cdot \vec{b}}
\int^{\infty}_{\infty} dz_1 \int^{\infty}_{z_1} dz_2
\int_{x_{\Pomeron}^{\rm min}}^{0.1} dx_{\Pomeron} \,\rho_A(b,z_1) \rho_A(b,z_2)
\nonumber\\
& \times k_{\eta} \, e^{-i m_N z_2 (x_{\Pomeron}-2\xi_N)+i m_N z_1 x_{\Pomeron}}
\phi_{\Pomeron /N}(x_{\Pomeron})
\phi_{\Pomeron /N}(x_{\Pomeron}-2\xi_N) \Big\}\frac{1}{x_{\Pomeron}}
{\cal H}_{\Pomeron}(\xi_{\Pomeron},t_{\rm min},Q^2) \,.
\label{eq:graph_b_4}
\end{align}

To the leading twist accuracy and to the leading order in the strong coupling constant,
${\cal H}_{\Pomeron}$ can be expressed in terms of the GPD of the Pomeron,
$H_{\Pomeron}$, as
\begin{equation}
{\cal H}_{\Pomeron}(\xi_{\Pomeron},t)=\int^1_{-1} dx^{\prime} H_{\Pomeron}(x^{\prime},\xi_{\Pomeron},t)
\left(\frac{1}{x^{\prime}-\xi_{\Pomeron}+i \epsilon}+\frac{1}{x^{\prime}+\xi_{\Pomeron}-i \epsilon} \right) \,.
\label{eq:Pomeron_GPD}
\end{equation}
Using the same argument that led to Eq.~(\ref{eq:proportionality}), we find that
\begin{equation}
\frac{x^{\prime}}{\xi_{\Pomeron}}=\frac{x}{\xi_A} =\frac{x_N}{\xi_N} \,,
\label{eq:proportionality_Pomeron}
\end{equation}
where $x$ parametrizes the interacting quark LC fractions in the graph in Fig.~\ref{fig:conversion}(a). Those fractions are equal to
$x+\xi_A=(\xi_A/\xi_{\Pomeron})(x^{\prime}+\xi_{\Pomeron})$ and 
$x-\xi_A=(\xi_A/\xi_{\Pomeron})(x^{\prime}-\xi_{\Pomeron})$,
 respectively.
Since $|x^{\prime}| \leq 1$, we find that $|x| \leq \xi_A/\xi_{\Pomeron}$.
Thus, substituting Eq.~(\ref{eq:graph_b_4}) into the first line of Eq.~(\ref{eq:CFF}), changing
the integration variable from $x$ to $x^{\prime}$ according to Eq.~(\ref{eq:proportionality_Pomeron}), 
recalling Eq.~(\ref{eq:Pomeron_GPD}), and noticing that the ensuing relation holds not only for the 
DVCS amplitude written to the leading order in the strong coupling constant, but also for individual
parton flavors, we obtain the contribution of the graph in Fig.~\ref{fig:DVCS_NS}(b) 
to the nuclear
GPD of flavor $j$, $H_{A}^{j(b)}$, as
\begin{align}
H_{A}^{j(b)}(x,\xi_A,t,Q^2)&=-
\frac{A(A-1)}{2} \frac{\xi_N}{\xi_A}
\, 16 \pi B_{\rm diff} \Re e \Big\{
 \int d^2 \vec{b}\, e^{i \vec{\Delta}_{\perp} \cdot \vec{b}}
\int^{\infty}_{\infty} dz_1 \int^{\infty}_{z_1} dz_2
\int_{x_{\Pomeron}^{\rm min}}^{0.1} dx_{\Pomeron} 
\nonumber\\
& \times \rho_A(b,z_1) \rho_A(b,z_2)
\, k_{\eta} \, e^{-i m_N z_2 (x_{\Pomeron}-2\xi_N)+i m_N z_1 x_{\Pomeron}}
\nonumber\\
& \times
\phi_{\Pomeron /N}(x_{\Pomeron})
\phi_{\Pomeron /N}(x_{\Pomeron}-2\xi_N)\Big\} \frac{1}{x_{\Pomeron}}
H^j_{\Pomeron}(\frac{\xi_{\Pomeron}}{\xi_N} x_N,\xi_{\Pomeron},t_{\rm min},Q^2) \,.
\label{eq:graph_b_5}
\end{align}
The GPD of the Pomeron, $H_{\Pomeron}^j$, 
is modeled by using the PDFs of the Pomeron, $f_{j/\Pomeron}$.
In our numerical analysis, we used the model of GPDs in which it is assumed that
the effect of skewedness in GPDs can be neglected at the initial evolution 
scale. This model corresponds to the double 
distribution model~\cite{Radyushkin:1997ki} with a $\delta$-function-like profile~\cite{Belitsky:2001ns}.
The details are given in Sec.~\ref{sec:results}

\subsection{Quasi-eikonal approximation for multiple rescatterings
 and the final expression for nuclear PDFs}
\label{subsec:quasi-eikonal}

To evaluate the contribution of the graph in Fig.~\ref{fig:DVCS_NS}(c),
we use the following high-energy (small $x_B$) space-time development of the process.
The virtual photon diffractively interacts with nucleon 1 and produces a certain 
diffractive state $X$ characterized by $x_{\Pomeron}$ (diffractive mass $M_X$). 
The produced state is then assumed to elastically scatter on $A-2$ nucleons of the target. Finally, the last interaction of the state $X$ with nucleon 2 produces the final real photon.
This picture of multiple rescattering at high-energy corresponds to the quasi-eikonal
approximation for the graph in Fig.~\ref{fig:DVCS_NS}(c) and higher rescattering terms.
The quasi-eikonal approximation
was used in the evaluation of nuclear PDFs in the framework of the leading twist theory of
nuclear shadowing~\cite{Frankfurt:1998ym,Frankfurt:2002kd,Frankfurt:2003zd} and in the evaluation of the DVCS amplitude on nuclei in the framework of Generalized vector meson dominance 
model~\cite{Goeke:2008rn}.

\begin{figure}[h]
\begin{center}
\epsfig{file=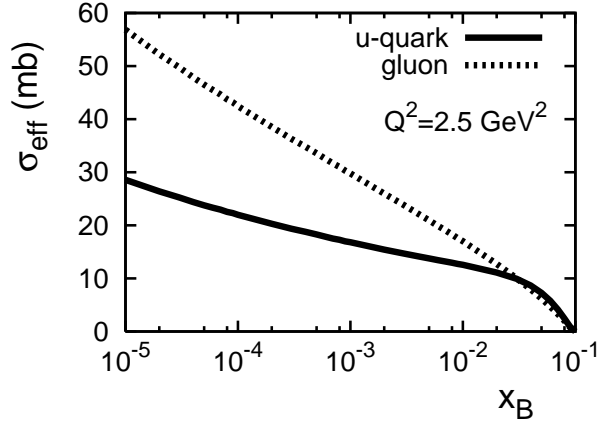,scale=1.3}
\caption{The effective cross section $\sigma_{\rm eff}^j$ [see Eq.~(\ref{eq:sigma_eff})]
for the $\bar{u}$ quarks and gluons as a function of Bjorken $x_B$ and at fixed
$Q^2=2.5$ GeV$^2$.
}
\label{fig:sigma_eff}
\end{center}
\end{figure}

Within the quasi-eikonal approximation, the multiple interactions can be summed and can be cast in the form of the eikonal attenuation factor, $T$,
\begin{equation}
T=e^{-\frac{A}{2} (1-i \eta) \sigma_{\rm eff}^j(x_B,Q^2) \int^{z_2}_{z_1}dz^{\prime} \rho_A(\vec{b},z^{\prime})} \,,
\label{eq:attenuation}
\end{equation}
 where 
$\sigma_{\rm eff}^j$ is the effective cross section, which determines the strength 
of the rescattering of the state $X$ off the nucleons. This cross section is defined
as~\cite{Frankfurt:1998ym,Frankfurt:2002kd,Frankfurt:2003zd}
\begin{equation}
\sigma_{\rm eff}^j(x_B,Q^2)=\frac{16 \pi B_{\rm diff}}{(1+\eta^2) x_B f_{j/N}(x_B,Q^2)}
\int^{0.1}_{x_B} dx_{\Pomeron} \beta f_{\Pomeron/p}(x_{\Pomeron}) f_{j/\Pomeron}(\beta,Q^2) \,,
\label{eq:sigma_eff}
\end{equation}
where $f_{j/N}$ is the usual parton PDF of the nucleon.
For a given flavor $j$, $\sigma_{\rm eff}^j$ is proportional to the probability of diffraction  relative to the total probability of the interaction. 
As an example, we present $\sigma_{\rm eff}^j$ as a function of $x_B$ at fixed
$Q^2=2.5$ GeV$^2$ for the $\bar{u}$ quark and gluon flavors in Fig.~\ref{fig:sigma_eff}.

Thus, collecting all contributions to the nuclear GPD $H_A^j$, 
\begin{equation}
H_A^j(x,\xi_A,t,Q^2)=H_A^{j(a)}+H_A^{j(b)}+H_A^{j(c)}+\dots \,,
\label{eq:sum}
\end{equation}
we obtain our final expression for flavor $j$ GPD of a heavy spinless nucleus:
\begin{align}
H_{A}^{j}(x,\xi_A,t,&Q^2)=\frac{\xi_N}{\xi_A}F_A(t) \sum_N H_N^j(x_N,\xi_N,t,Q^2) \nonumber\\
&-
\frac{A(A-1)}{2} \frac{\xi_N}{\xi_A}
\, 16 \pi B_{\rm diff} \,\Re e\Big\{
 \int d^2 \vec{b}\, e^{i \vec{\Delta}_{\perp} \cdot \vec{b}}
\int^{\infty}_{\infty} dz_1 \int^{\infty}_{z_1} dz_2
\int_{x_{\Pomeron}^{\rm min}}^{0.1} dx_{\Pomeron} 
\nonumber\\
& \times \rho_A(b,z_1) \rho_A(b,z_2)
\, k_{\eta} \,e^{-i m_N z_2 (x_{\Pomeron}-2\xi_N)+i m_N z_1 x_{\Pomeron}}
e^{-\frac{A}{2} (1-i \eta) \sigma_{\rm eff}^j(x_B,Q^2) \int^{z_2}_{z_1}dz^{\prime} \rho_A(\vec{b},z^{\prime})}
\nonumber\\
& \times
\phi_{\Pomeron /N}(x_{\Pomeron})
\phi_{\Pomeron /N}(x_{\Pomeron}-2\xi_N)\Big\} \frac{1}{x_{\Pomeron}} H^j_{\Pomeron}(\frac{\xi_{\Pomeron}}{\xi_N} x_N,\xi_{\Pomeron},t_{\rm min},Q^2) \,.
\label{eq:final_0}
\end{align}
For practical applications and for a comparison to the case of a free nucleon, it is convenient to simultaneously rescale the LC fraction $x$ and the nuclear GPDs in the left-hand side of Eq.~(\ref{eq:final_0}):
\begin{equation}
H_{A}^{j}(x,\xi_A,t,Q^2) \to \frac{\xi_N}{\xi_A} H_{A}^{j}(x_N,\xi_A,t,Q^2) \,,
\end{equation}
(where the rescaling of the nuclear GPD is necessary to preserve sum rules involving the nuclear GPD).
Then, our master equation for the nuclear GPD becomes
\begin{align}
H_{A}^{j}(x_N,\xi_A,t,&Q^2)=F_A(t) \sum_N H_N^j(x_N,\xi_N,t,Q^2) \nonumber\\
&-
\frac{A(A-1)}{2} 
\, 16 \pi B_{\rm diff} \,\Re e \Big\{
 \int d^2 \vec{b}\, e^{i \vec{\Delta}_{\perp} \cdot \vec{b}}
\int^{\infty}_{\infty} dz_1 \int^{\infty}_{z_1} dz_2
\int_{x_{\Pomeron}^{\rm min}}^{0.1} dx_{\Pomeron} 
\nonumber\\
& \times \rho_A(b,z_1) \rho_A(b,z_2)
\, k_{\eta} \,e^{-i m_N z_2 (x_{\Pomeron}-2\xi_N)+i m_N z_1 x_{\Pomeron}}
e^{-\frac{A}{2} (1-i \eta) \sigma_{\rm eff}^j(x_B,Q^2) \int^{z_2}_{z_1}dz^{\prime} \rho_A(\vec{b},z^{\prime})}
\nonumber\\
& \times
\phi_{\Pomeron /N}(x_{\Pomeron})
\phi_{\Pomeron /N}(x_{\Pomeron}-2\xi_N) \Big\}
\frac{1}{x_{\Pomeron}}
 H^j_{\Pomeron}(\frac{\xi_{\Pomeron}}{\xi_N} x_N,\xi_{\Pomeron},t_{\rm min},Q^2) \,.
\label{eq:final}
\end{align}
As we explained above, we neglected the Fermi motion effect in the first term in 
Eq.~(\ref{eq:final}). If necessary, the Fermi motion effect can be restored 
by replacing the first term in Eq.~(\ref{eq:final}) by the right-hand side of
Eq.~(\ref{eq:graph_a_7}).

\section{Nuclear GPDs in the $\xi_A \to 0$ limit and the spacial image
of nuclear shadowing}
\label{sec:image}

In the forward limit, nuclear GPDs reduce to nuclear PDFs,
\begin{equation}
H_{A}^{j}(x,0,0,Q^2)=f_{j/A}(x,Q^2) \,.
\label{eq:fl1}
\end{equation}
Taking the $\xi_A=t=0$ limit in Eq.~(\ref{eq:final}), we  obtain 
\begin{align}
H_{A}^{j}(x_B,0,0,Q^2)&=
\sum_N f_{j/N}(x_B,Q^2) \nonumber\\
&-
\frac{A(A-1)}{2} 
\, 16 \pi B_{\rm diff}\,\Re e\Big\{
 \int d^2 \vec{b}\, 
\int^{\infty}_{\infty} dz_1 \int^{\infty}_{z_1} dz_2
\int_{x_B}^{0.1} dx_{\Pomeron} 
\nonumber\\
& \times \rho_A(b,z_1) \rho_A(b,z_2)
\, k_{\eta} \,e^{i m_N x_{\Pomeron}(z_1-z_2)}
e^{-\frac{A}{2} (1-i \eta) \sigma_{\rm eff}^j(x_B,Q^2) \int^{z_2}_{z_1}dz^{\prime} \rho_A(\vec{b},z^{\prime})}
\nonumber\\
& \times f_{\Pomeron /p}(x_{\Pomeron})\frac{1}{x_{\Pomeron}}
 f_{j/\Pomeron}(\beta=\frac{x_B}{x_{\Pomeron}},Q^2) \Big\} \,.
\label{eq:fl2}
\end{align}
Here we used the fact that, in the $\xi_A \to 0$ limit, $\xi_N,\xi_{\Pomeron},t_{\rm min} \to 0$
and $\xi_{\Pomeron} x_N /\xi_N=x^{\prime}\to \beta=x_B/x_{\Pomeron}$.
The obtained expression for the nuclear PDF $f_{j/A}$ as a forward limit of the nuclear GPD
coincides with the direct calculation of $f_{j/A}$ in the framework of the leading twist
theory of nuclear shadowing~\cite{Frankfurt:1998ym,Frankfurt:2002kd,Frankfurt:2003zd};
that is, our master equation [Eq.~(\ref{eq:final})] has the correct (consistent) forward limit.

Next let us consider the $\xi_A \to 0$ limit (i.e., the limit when the momentum
transfer $t$ is purely transverse, $t=-\Delta^2_{\perp}$).
Taking the $\xi_A \to 0$ limit in Eq.~(\ref{eq:final}), we obtain
\begin{align}
H_{A}^{j}(x_N,0,t,Q^2)&=F_A(t) \sum_N H_N^j(x_N,0,t,Q^2) \nonumber\\
&-
\frac{A(A-1)}{2} 
\, 16 \pi B_{\rm diff} \, \Re e \Big\{
 \int d^2 \vec{b}\, e^{i \vec{\Delta}_{\perp} \cdot \vec{b}}
\int^{\infty}_{\infty} dz_1 \int^{\infty}_{z_1} dz_2
\int_{x_N}^{0.1} dx_{\Pomeron} 
\nonumber\\
& \times \rho_A(b,z_1) \rho_A(b,z_2)
\, k_{\eta} \,e^{i m_N x_{\Pomeron}(z_1-z_2)}
e^{-\frac{A}{2} (1-i \eta) \sigma_{\rm eff}^j(x_B,Q^2) \int^{z_2}_{z_1}dz^{\prime} \rho_A(\vec{b},z^{\prime})}
\nonumber\\
& \times
f_{\Pomeron /p}(x_{\Pomeron}) \frac{1}{x_{\Pomeron}} f_{j/\Pomeron}\big(\frac{x_N}{x_{\Pomeron}},Q^2\big)
 \Big\} \,.
\label{eq:xiAlimit}
\end{align}
Again, we used the fact that, in the $\xi_A \to 0$ limit, $\xi_N,\xi_{\Pomeron},t_{\rm min} \to 0$, $\xi_{\Pomeron} x_N /\xi_N \to x_N/x_{\Pomeron}$,
and $H^j_{\Pomeron}(\frac{\xi_{\Pomeron}}{\xi_N} x_N,\xi_{\Pomeron},t_{\rm min},Q^2) \to 
f_{j/\Pomeron}(x_N/x_{\Pomeron},Q^2)$.
Note also the lower limit of integration over $x_{\Pomeron}$,  
$x_{\Pomeron}^{\rm min}=x_N$.
Since the $t$ dependence of the nuclear form factor, $F_A(t)$, is much faster than that
of the nucleon GPD $H_N^j(x_N,0,t,Q^2)$, the latter can be evaluated at $t=0$ (i.e., in the
forward limit). Then, Eq.~(\ref{eq:xiAlimit}) becomes
\begin{align}
H_{A}^{j}(x_N,0,t,Q^2)&=F_A(t) \sum_N f_{j/N}(x_N,Q^2) \nonumber\\
&-
\frac{A(A-1)}{2} 
\, 16 \pi B_{\rm diff} \, \Re e \Big\{
 \int d^2 \vec{b}\, e^{i \vec{\Delta}_{\perp} \cdot \vec{b}}
\int^{\infty}_{\infty} dz_1 \int^{\infty}_{z_1} dz_2
\int_{x_N}^{0.1} dx_{\Pomeron} 
\nonumber\\
& \times \rho_A(b,z_1) \rho_A(b,z_2)
\, k_{\eta} \,e^{i m_N x_{\Pomeron}(z_1-z_2)}
e^{-\frac{A}{2} (1-i \eta) \sigma_{\rm eff}^j(x_B,Q^2) \int^{z_2}_{z_1}dz^{\prime} \rho_A(\vec{b},z^{\prime})}
\nonumber\\
& \times
f_{\Pomeron /p}(x_{\Pomeron}) \frac{1}{x_{\Pomeron}} f_{j/\Pomeron}\big(\frac{x_N}{x_{\Pomeron}},Q^2\big)
\Big\} \,.
\label{eq:xiAlimit2}
\end{align}

In the case of nucleon GPDs, the interpretation of GPDs in the $\xi \to 0$ limit
is given in the impact parameter representation, where the GPDs have the meaning
of the probability densities~\cite{Burkardt:2002hr}.
We shall also analyse our nuclear GPDs in the $\xi_A \to 0$ limit in the impact
parameter space. To this end, we introduce the nuclear GPD in the impact parameter space,
\begin{equation}
H_{A}^{j}(x,0,\vec{b},Q^2)=\int \frac{d^2 \vec{\Delta}_{\perp}}{(2 \pi)^2} e^{-i \vec{\Delta}_{\perp} \cdot \vec{b}}\,H_{A}^{j}(x,0,t=-\Delta^2_{\perp},Q^2) \,.
\label{eq:impact}
\end{equation}
The Fourier transform of Eq.~(\ref{eq:xiAlimit2}) gives
\begin{align}
H_{A}^{j}(x_N,0,\vec{b},Q^2)&=T_A(b) \sum_N f_{j/N}(x_N,Q^2) \nonumber\\
&-\frac{A(A-1)}{2}  \, 16 \pi B_{\rm diff} \, \Re e \Big\{
\int^{\infty}_{\infty} dz_1 \int^{\infty}_{z_1} dz_2
\int_{x_N}^{0.1} dx_{\Pomeron} 
\nonumber\\
& \times \rho_A(b,z_1) \rho_A(b,z_2)
\, e^{i m_N x_{\Pomeron}(z_1-z_2)}
e^{-\frac{A}{2} (1-i \eta) \sigma_{\rm eff}^j(x_B,Q^2) \int^{z_2}_{z_1}dz^{\prime} \rho_A(\vec{b},z^{\prime})}
\nonumber\\
& \times
f_{\Pomeron /p}(x_{\Pomeron}) \frac{1}{x_{\Pomeron}} f_{j/\Pomeron}\big(\frac{x_N}{x_{\Pomeron}},Q^2\big)
 \Big\} \,,
\label{eq:impact2}
\end{align}
where $T_A(b)=\int dz \rho_A(b,z)$ and $\rho_A(b,z)$ is the nuclear density
[see Eq.~(\ref{eq:wf_normalization_coordinate})].
It is important to note that the nuclear GPD $H_{A}^{j}(x_N,0,\vec{b},Q^2)$
given by Eq.~(\ref{eq:impact2}) is nothing else but the impact-parameter-dependent nuclear PDF introduced and discussed in the framework of the leading twist nuclear shadowing~\cite{Frankfurt:1998ym,Frankfurt:2002kd,Frankfurt:2003zd}.

\begin{figure}[h]
\begin{center}
\epsfig{file=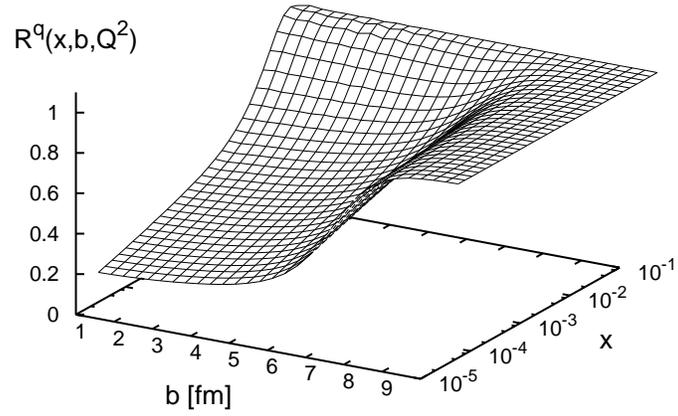,scale=1.0}
\epsfig{file=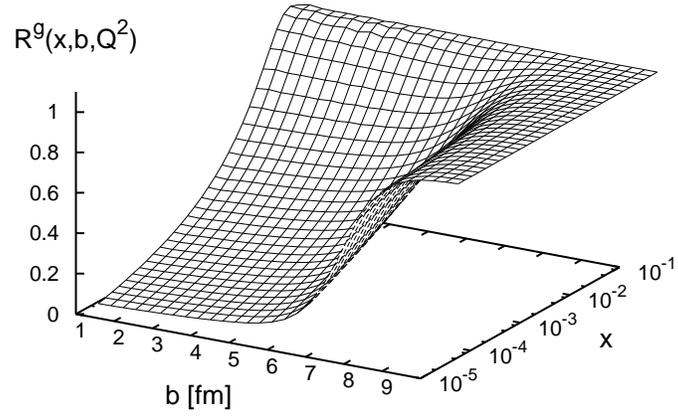,scale=1.0}
\caption{Impact parameter dependence of nuclear shadowing for $^{208}$Pb. 
The graphs show
the ratio
$R^j(x_N,b,Q^2)$ of Eq.~(\ref{eq:R}) as a function of the LC fraction
$x_N$ and the impact parameter $b$ at fixed $Q^2=2.5$ GeV$^2$.
The top panel corresponds to $\bar{u}$ quarks;
the bottom panel corresponds to gluons.}
\label{fig:R}
\end{center}
\end{figure}

In Eq.~(\ref{eq:impact2}), the first term is the Born approximation to $H^j_A$ corresponding
to the graph in Fig.~\ref{fig:DVCS_NS}(a); the second term is the nuclear shadowing 
correction corresponding to the graphs in  Figs.~\ref{fig:DVCS_NS}(b) and  \ref{fig:DVCS_NS}(c) and to higher rescattering terms not shown in 
Fig.~\ref{fig:DVCS_NS}. We quantify the magnitude of the nuclear shadowing correction
by considering the ratio
\begin{equation}
R^j(x_N,b,Q^2)=\frac{H_{A}^{j}(x_N,0,\vec{b},Q^2)}{T_A(b) \sum_N f_{j/N}(x_N,Q^2)} \,,
\label{eq:R}
\end{equation}
where the numerator is given by Eq.~(\ref{eq:impact2}). 
In the absence of nuclear shadowing,
$R^j(x_N,b,Q^2)=1$.
The ratio $R^j(x_N,b,Q^2)$ for the nucleus of $^{208}$Pb
as a function of $x_N$ and $b$ at fixed $Q^2=2.5$ GeV$^2$ is presented in 
Fig.~\ref{fig:R}. In the figure, the top panel corresponds to $\bar{u}$ quarks;
the bottom panel corresponds to gluons.

Essentially, Fig.~\ref{fig:R} presents the impact parameter dependence
of nuclear shadowing, or the spacial image of nuclear shadowing.
Several features of Fig.~\ref{fig:R} deserve mentioning. First,
the amount of nuclear shadowing [the deviation of $R^j(x_N,b,Q^2)$ from unity]
increases as one decreases $x_N$ and $b$. Second, nuclear shadowing for gluons
is larger than for quarks. For instance, at $x_N=10^{-5}$ and at $b=0$,
$R^g=0.073$, but $R^q=0.23$. Third, nuclear shadowing induces non-trivial correlations between $x_N$ and $b$ in the nuclear GPD $H_{A}^{j}(x,0,\vec{b},Q^2)$, even if such correlations are absent in the free nucleon GPD.
[In Eq.~(\ref{eq:impact2}) we neglected the $x_N$-$b$ correlations in the nucleon 
GPDs by neglecting the $t$ dependence of $H_N^j(x_N,0,t,Q^2)$.]
In this respect, the spacial image of nuclear GPDs at small $x_N$ is very different
from the case of the free nucleon: Whereas the free nucleon GPDs become independent
of $b$ in the $x_N \to 0$ limit~\cite{Burkardt:2002hr}, the suppression of nuclear
GPDs by nuclear shadowing is strongly correlated with the impact parameter 
$b$.

\section{Nuclear shadowing and predictions for nuclear DVCS observables}
\label{sec:results}

It is convenient to quantify the amount of nuclear shadowing in our master expression 
for the nuclear GPD of a heavy nucleus [Eq.~(\ref{eq:final})]
in terms of the $R^j(x_N,\xi_N,t,Q^2)$ ratio, which we define as
\begin{align}
R^j(x_N,&\xi_N,t,Q^2) \equiv \frac{H_{A}^{j}(x_N,\xi_A,t,Q^2)}{F_A(t)\sum_N H_N^j(x_N,\xi_N,t,Q^2)} \nonumber\\
&=1-\frac{A(A-1)}{2} 
\, 16 \pi B_{\rm diff} \,\Re e \Big\{
 \int d^2 \vec{b}\, e^{i \vec{\Delta}_{\perp} \cdot \vec{b}}
\int^{\infty}_{\infty} dz_1 \int^{\infty}_{z_1} dz_2
\int_{x_{\Pomeron}^{\rm min}}^{0.1} dx_{\Pomeron} 
\nonumber\\
& \times \rho_A(b,z_1) \rho_A(b,z_2)
\, k_{\eta} \,e^{-i m_N z_2 (x_{\Pomeron}-2\xi_N)+i m_N z_1 x_{\Pomeron}}
e^{-\frac{A}{2} (1-i \eta) \sigma_{\rm eff}^j(x_B,Q^2) \int^{z_2}_{z_1}dz^{\prime} \rho_A(\vec{b},z^{\prime})}
\nonumber\\
& \times
\phi_{\Pomeron /N}(x_{\Pomeron})
\phi_{\Pomeron /N}(x_{\Pomeron}-2\xi_N) \Big\}
\frac{1}{x_{\Pomeron}}
 H^j_{\Pomeron}(\frac{\xi_{\Pomeron}}{\xi_N} x_N,\xi_{\Pomeron},t_{\rm min},Q^2)
\nonumber \\
&\Big/ \left(F_A(t) \sum_NH_N^j(x_N,\xi_N,t,Q^2)\right) \,.
\label{eq:R_gpd}
\end{align}
In the absence of nuclear shadowing (and the Fermi motion effect), $R^j(x_N,\xi_N,t,Q^2)=1$.
The ratio $R^j(x_N,\xi_N,t,Q^2)$ is a generalization and a Fourier transform of the ratio $R^j(x_N,b,Q^2)$ of Eq.~(\ref{eq:R}).

At high energies, scattering amplitudes are predominantly imaginary.
As follows from Eq.~(\ref{eq:CFF}), to the leading twist accuracy and to the leading order in $\alpha_s$,
the imaginary part of the 
DVCS amplitude (the CFF) reads
\begin{equation}
\Im m {\cal H}_A(\xi_A,t,Q^2)=-\pi H_{A}(\xi_A,\xi_A,t,Q^2) \,,
\label{eq:singlet1}
\end{equation}
where
\begin{equation}
H_{A}(\xi_A,\xi_A,t)=\sum_{q,\bar{q}} e_q^2 H_{A}^{q}(\xi_A,\xi_A,t) \,.
\label{eq:singlet2}
\end{equation}
Therefore, in our numerical analysis that follows, we shall present
our predictions for  $R^j(\xi_N,\xi_N,t,Q^2)$.

In our numerical analysis, we use the model of GPDs of the free nucleon 
and the Pomeron, in which it is  assumed that the effect of skewedness in 
GPDs can be neglected at the initial QCD evolution scale ($Q^2_0=2.5$ 
GeV$^2$ in our case). Then, in the $x_N=\xi_N$ case of interest, one has
\begin{eqnarray}
H_N^j(\xi_N,\xi_N,Q^2_0)&=&f_{j/N}(\xi_N,Q^2_0) \nonumber\\
H_{\Pomeron}^j(\xi_{\Pomeron},\xi_{\Pomeron},Q^2_0)&=&
f_{j/\Pomeron}(\xi_{\Pomeron},Q^2_0)=f_{j/\Pomeron}(\frac{\xi_N}{x_{\Pomeron}},Q^2_0)\,.
\label{eq:model1}
\end{eqnarray}
This model corresponds to the double 
distribution parameterization of GPDs~\cite{Radyushkin:1997ki} with a 
$\delta$-function-like profile~\cite{Belitsky:2001ns}; we shall refer to this model of the GPDs
as the forward-like model. Note that the suggestion that the GPDs at small $x_B$ and at
the low input scale $Q_0^2$ can be well approximated by the usual forward PDFs was
first proposed in Ref.~\cite{Frankfurt:1997ha}. 

It is very important to point out that the recent analysis of the high-energy 
HERA data on DVCS on
the proton unambiguously indicated that the description of the
data at the leading order accuracy requires almost no skewedness effect in the 
input GPDs~\cite{Kumericki:2008sb}. 
This clearly favors the forward-like model of the PDFs over other
small-$x_B$ parameterizations (see, e.g., Ref.~\cite{Guzey:2008ys}).

\begin{figure}[t]
\begin{center}
\epsfig{file=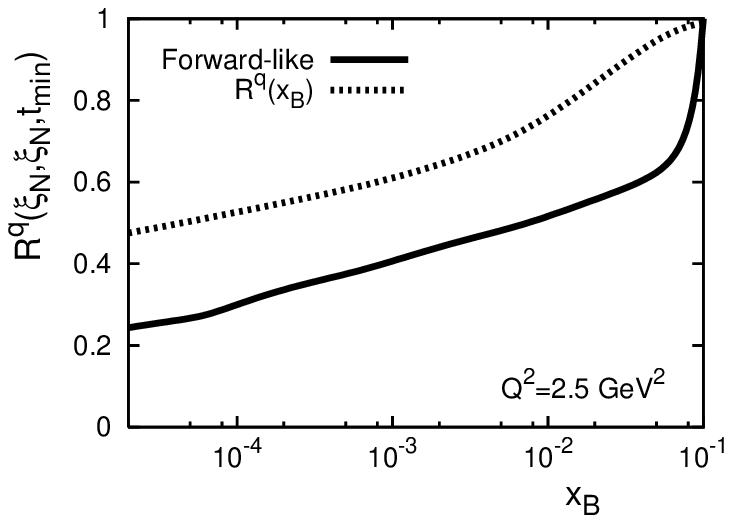,scale=1.05}
\epsfig{file=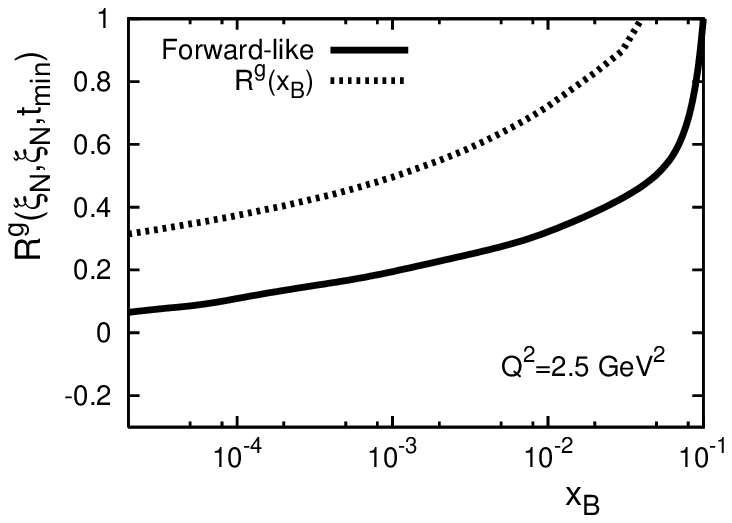,scale=1.05}
\caption{Nuclear shadowing for the DVCS amplitude for $^{208}$Pb at $\vec{\Delta}_{\perp}=0$. The plots show
the ratio $R^j(\xi_N,\xi_N,t,Q^2)$  of Eq.~(\ref{eq:R_gpd}) as a function of $x_B$ at fixed $Q^2_0=2.5$ GeV$^2$ for the forward-like model of GPDs (solid curves). 
For comparison, the ratio of the usual nuclear to nucleon PDFs, 
$R^j(x_B)=f_{j/A}(x_B,Q_0^2)/[A f_{j/N}(x_B,Q_0^2)]$, is given by the dotted curves.
The left panel corresponds to $\bar{u}$ quarks; the right panel corresponds to gluons.
}
\label{fig:R_xixi}
\end{center}
\end{figure}

Let us first examine the $R^j(\xi_N,\xi_N,t,Q^2)$ ratio of Eq.~(\ref{eq:R_gpd})
in the situation
when the momentum transfer $t$ is purely longitudinal,
$\vec{\Delta}_{\perp}=0$ and $t=t_{\rm min} \approx -4 \xi_N^2 m_N^2$.
Figure~\ref{fig:R_xixi} presents $R^j(\xi_N,\xi_N,t,Q^2)$ for $^{208}$Pb 
as a function of Bjorken $x_B$ at fixed $Q^2_0=2.5$ GeV$^2$ 
(solid curves). 
Also, for comparison with nuclear shadowing in usual nuclear PDFs, we present
the ratio $R^j(x_B)=f_{j/A}(x_B,Q_0^2)/[A f_{j/N}(x_B,Q_0^2)]$ by the dotted curves~\cite{Frankfurt:2003zd}.
The left panel corresponds to $\bar{u}$ quarks;
the right panel corresponds to gluons.

As one can see from Fig.~\ref{fig:R_xixi}, 
the suppression of $R^j(\xi_N,\xi_N,t,Q_0^2)$ by nuclear shadowing 
is very large and it is larger than the suppression of $R^j(x_B,Q_0^2)$ in the forward 
case. This is one of new results of this work and it comes from our model for 
graph  in Fig.~\ref{fig:graph_b}(b). In particular, we assumed that the matrix element
\begin{equation}
\langle p_1^{\prime} p_2^{\prime}| T\{J^{\mu}(x) J^{\nu}(0)\} |p_1 p_2 \rangle
\propto \phi_{\Pomeron /N}(x_{\Pomeron})\phi_{\Pomeron /N}(x_{\Pomeron}-2\xi_N) \,, 
\label{eq:key}
\end{equation}
which leads to the dynamical enhancement of nuclear shadowing because 
$\phi_{\Pomeron /N}(x_{\Pomeron}-2\xi_N) \gg \phi_{\Pomeron /N}(x_{\Pomeron})$ for $x_{\Pomeron}$ close to $2 \xi_N$.

We stress that our results presented in Fig.~\ref{fig:R_xixi} have an
exploratory character and are subject of significant theoretical uncertainties, 
which include our modeling of the  $\langle p_1^{\prime} p_2^{\prime}| T\{J^{\mu}(x) J^{\nu}(0)\} |p_1 p_2 \rangle$ matrix element, the choice of the model 
for the nucleon and Pomeron GPDs, and the extrapolation of the fits for diffractive PDFs $f^j_{\Pomeron}$ to unmeasured kinematic regions. 

We also mention that the rapid approach of $R^j(\xi_N,\xi_N,t,Q_0^2)$ to unity
as $x_B \to 0.1$ is driven both by the decrease of the nuclear shadowing  
term and by the decrease of the Born term driven by the nuclear form factor at
$t=t_{\rm min} \approx - x_B^2 m_N^2$, $F_A(t_{\rm min})$.

\begin{figure}[t]
\begin{center}
\epsfig{file=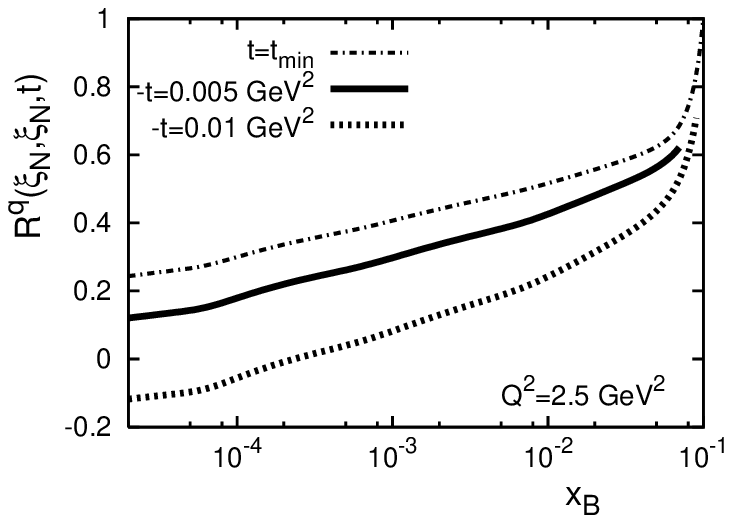,scale=1.05}
\epsfig{file=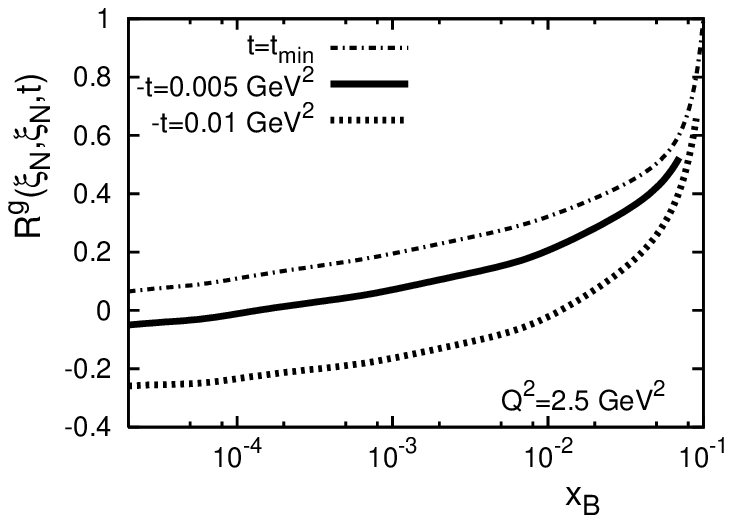,scale=1.05}
\caption{Nuclear shadowing for the DVCS amplitude for $^{208}$Pb at fixed $t$.
The plots show
the ratio $R^j(\xi_N,\xi_N,t,Q^2)$ of Eq.~(\ref{eq:R_gpd}) as a function of $x_B$ at fixed $Q^2_0=2.5$ GeV$^2$.
The left panel corresponds to $\bar{u}$ quarks; the right panel 
corresponds to gluons.
The solid curves correspond to $t=-0.005$ GeV$^2$; the dotted curves correspond to
$t=-0.01$ GeV$^2$.
For comparison, the ratio $R^j(\xi_N,\xi_N,t_{\rm min},Q^2)$ at $t=t_{\rm min}$
is given by the dot-dashed curves.
}
\label{fig:R_xixi_fixed_t}
\end{center}
\end{figure}

Next we examine the ratio $R^j(\xi_N,\xi_N,t,Q^2)$ at fixed $t$ as a function of $x_B$.
In this case, the transverse momentum transfer is no longer vanishing:
$|\vec{\Delta}_{\perp}|^2 \approx -4 \xi_N^2 m_N^2 -t$.
Our results are presented in Fig.~\ref{fig:R_xixi_fixed_t}.
The left panel corresponds to $\bar{u}$ quarks; the right panel 
correspond to gluons.
The solid curves correspond to $t=-0.005$ GeV$^2$; the dotted curves correspond to
$t=-0.01$ GeV$^2$.
For comparison, the ratio $R^j(\xi_N,\xi_N,t_{\rm min},Q^2)$ at $t=t_{\rm min}$
is given by the dot-dashed curves (the same curves as in Fig.~\ref{fig:R_xixi}).

As one can see from Fig.~\ref{fig:R_xixi_fixed_t}, the effect of nuclear shadowing
[the deviation of $R^j(\xi_N,\xi_N,t,Q^2)$ from unity at small $x_B$] increases with
increasing $|t|$. This is a natural consequence of the fact the Born term, whose
$t$ dependence is given by $F_A(t)$, decreases with increasing $|t|$ faster than
the shadowing correction term.

Next we turn to a discussion of observables measured in DVCS.
In lepton-nucleus scattering, it is convenient and natural to use 
the invariant energy per nucleon. For our results presented in the following, 
this means that we replace $\xi_A \to \xi_N$ and assume that the invariant energy, $\sqrt{s}$, is given per nucleon. 
Results of high-energy DVCS measurements are usually presented in terms of the DVCS
cross section at the photon level, 
\begin{equation}
\frac{d \sigma_{\rm DVCS}}{dt}=\frac{\pi \alpha_{\rm em}^2 x_B^2}{Q^4}|{\cal A}_{\rm DVCS}(\xi_N,t,Q^2)|^2 \,,
\label{eq:sigma_dvcs}
\end{equation}
where $\alpha_{\rm em}$ is the fine-structure constant.
For the DVCS amplitude at high energies, we use the leading twist and leading order in $\alpha_s$ expression
[see Eqs.~(\ref{eq:singlet1}) and (\ref{eq:singlet2})],
\begin{equation}
|{\cal A}_{\rm DVCS}(\xi_N,t,Q^2)|^2 \approx |{\cal H}_A(\xi_N,t,Q^2)|^2 \approx \pi^2 (H_A(\xi_N,\xi_N,t,Q^2))^2 \,,
\end{equation}
where $H_A(\xi_N,\xi_N,t) =\sum_{q,\bar{q}} e_q^2 H_A^q(\xi_N,\xi_N,t)$ and
$H_A^q(\xi_N,\xi_N,t)$ are given by our master equation [Eq.~(\ref{eq:final})].
Since gluons enter the DVCS amplitude at the one-loop level, we do not use our 
results for the gluon nuclear GPD in our calculations presented in the following.
Note also that since we do our calculations at fixed $Q_0^2=2.5$ GeV$^2$, we use
four quark flavors.

The DVCS process competes with the purely electromagnetic Bethe-Heitler (BH) process.
The BH cross section at the photon level can be written in the following form
(see, e.g., Ref.~\cite{Belitsky:2001ns}):
\begin{equation}
\frac{d \sigma_{\rm BH}}{dt}=\frac{\pi \alpha_{\rm em}^2 }{4Q^2 t (1-y+y^2/2)}
\int^{2 \pi}_{0} \frac{d \phi}{2 \pi}\,\frac{1}{{\cal P}_1(\phi) {\cal P}_2(\phi)}
|{\cal A}_{\rm BH}(\xi_N,t,Q^2)|^2 \,,
\label{eq:sigma_bh}
\end{equation}
where $y$ is the fractional energy loss of the incoming lepton, $\phi$ is the angle between
the lepton and hadron scattering planes, ${\cal P}_1$ and ${\cal P}_2$ are proportional
to the lepton propagators, and 
$|{\cal A}_{\rm BH}(\xi_N,t,Q^2)|^2$ is the BH amplitude squared, which
 can be expressed in terms of its Fourier harmonics $c_n^{\rm BH}$~\cite{Belitsky:2001ns} as
\begin{equation}
|{\cal A}_{\rm BH}(\xi_N,t,Q^2,\phi)|^2=c_0^{\rm BH}+\sum_{n=1}^2 c_n^{\rm BH} \cos(n \phi) \,.
\label{eq:BH}
\end{equation}
The Fourier harmonics for a spinless target are given in Ref.~\cite{Belitsky:2000vk}.
For the case of a spinless nucleus,  
 $|{\cal A}_{\rm BH}(\xi_N,t,Q^2)|^2 \propto [F_A(t)]^2$;
see further details in Ref.~\cite{Guzey:2008th}.  

\begin{figure}[t]
\begin{center}
\epsfig{file=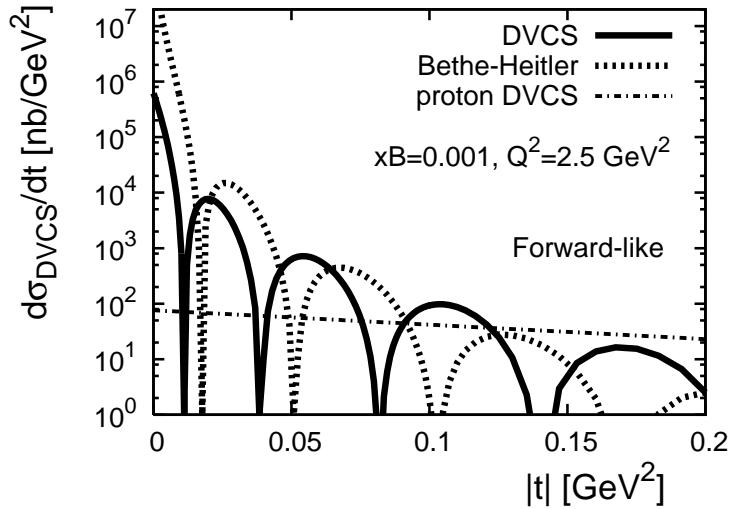,scale=1.3}
\caption{Nuclear DVCS and BH cross sections for $^{208}$Pb as a function of $|t|$ 
at fixed $Q_0^2=2.5$ GeV$^2$ and $x_B=0.001$.
For comparison, the DVCS cross section on the proton is given by the dot-dashed curves.
For the evaluation of the BH cross section, we used $y=0.31$ (see the text).
}
\label{fig:sigma_dvcs_tdep}
\end{center}
\end{figure}

Figure~\ref{fig:sigma_dvcs_tdep} presents our predictions for 
$d \sigma_{\rm DVCS}/dt$ and $d \sigma_{\rm BH}/dt$ for $^{208}$Pb as a function of $|t|$
at fixed $Q_0^2=2.5$ GeV$^2$ and $x_B=0.001$. In addition to the input discussed above, for the evaluation of the BH 
cross section, we used $y=0.31$, which corresponds to the highest among discussed energy 
options of the future Electron-Ion Collider (EIC), $E_{\rm lepton}=20$ GeV and $E_{\rm nucleus}=100$ GeV/nucleon~\cite{Deshpande:2005wd,eic}. Also, for comparison, we give the DVCS cross section on the
proton in the same kinematics (dot-dashed curves).

Several features of Fig.~\ref{fig:sigma_dvcs_tdep} deserve a discussion.
First, the $t$ dependence of the BH and DVCS cross sections repeats the pattern of 
$[F_A(t)]^2$ with several distinct minima. In the case of the DVCS cross section, 
the minima are shifted because of the presence of the shadowing
correction. 
Second, at small $|t|$, 
the BH cross section is much larger than the DVCS cross 
section owing to the enhancement by the $1/t$ kinematics factor [see Eq.~(\ref{eq:sigma_bh})].
As one increases $|t| > |t_{\rm min}|$, the two cross sections become compatible.
Moreover, near minima of the nuclear form factor, the BH cross section becomes very 
small and, hence, the process is dominated by the DVCS cross section.
{\it Therefore, the measurement of the $e A \to e \gamma A$ differential cross section at
the momentum transfer $t$ near the minima of the nuclear form factor
will provide a clean probe of nuclear shadowing in nuclear GPDs and nuclear DVCS
 owing to the suppressed
BH background and the suppressed unshadowed Born contribution to the DVCS
amplitude.}

\begin{figure}[t]
\begin{center} 
\epsfig{file=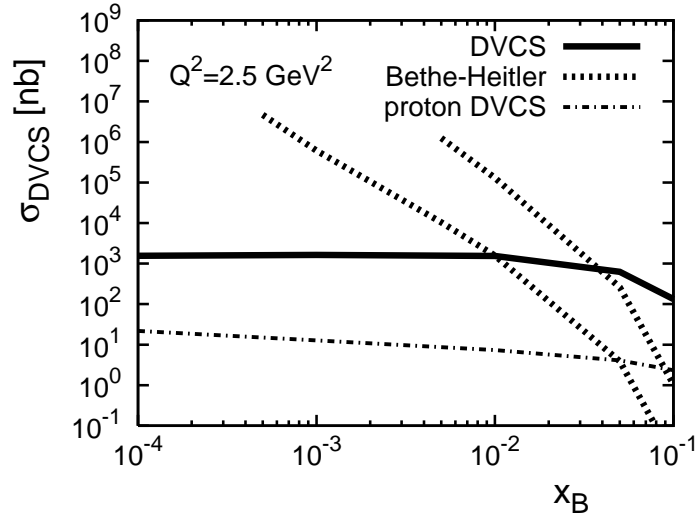,scale=1.3}
\caption{The $t$-integrated nuclear DVCS and BH cross sections 
for $^{208}$Pb as a function of $x_B$ at fixed $Q_0^2=2.5$ GeV$^2$.
For comparison, the DVCS cross section on the proton is given by the dot-dashed curves.
For the evaluation of the BH cross section, we used two energy settings:
$\sqrt{s}=32$ GeV (the upper dashed curve) and $\sqrt{s}=90$ GeV 
(the lower dashed curve)
(see the text).
}
\label{fig:sigma_dvcs_t_integrated}
\end{center}
\end{figure}

Next we study the $t$-integrated DVCS and BH cross sections at the photon level,
\begin{eqnarray}
\sigma_{\rm DVCS}&=&\int_{-1\ {\rm GeV}^2}^{t_{\rm min}} dt\, \frac{\sigma_{\rm DVCS}}{dt} \,,
\nonumber\\
\sigma_{\rm BH}&=&\int_{-1\ {\rm GeV}^2}^{t_{\rm min}} dt\, \frac{\sigma_{\rm BH}}{dt} \,.
\label{eq:t_integrated}
\end{eqnarray}
Figure~\ref{fig:sigma_dvcs_t_integrated} presents the $t$-integrated 
DVCS and BH cross sections for $^{208}$Pb as a function of $x_B$ 
at fixed $Q_0^2=2.5$ GeV$^2$. For comparison, the dot-dashed curve shows
the DVCS cross section on the proton in the same kinematics.
For the BH cross section, we give two curves, which correspond to two
different values of the c.m.~lepton-nucleus energy $\sqrt{s}$:
The upper curve corresponds to the low-energy option for the future EIC, $E_{\rm lepton}=5$ GeV and $E_{\rm nucleus}=50$ GeV/nucleon 
($\sqrt{s}=32$ GeV);
the lower curve corresponds to the high-energy option with 
$E_{\rm lepton}=20$ GeV and $E_{\rm nucleus}=100$ GeV/nucleon ($\sqrt{s}=90$ GeV)~\cite{Deshpande:2005wd,eic}.

As one see from Fig.~\ref{fig:sigma_dvcs_t_integrated}, in the discussed kinematics,
the BH cross section is much larger than the DVCS cross section for $x_B < 0.01$
for both considered high-energy options (lower BH curve) 
and for  $x_B < 0.05$ for the low-energy option (upper BH curve).
Therefore, as far as the $t$-integrated $e A \to e \gamma A$ cross section is concerned,
it appears rather challenging to extract a small DVCS signal on the background of the
dominant BH contribution. However, the high luminosity of the future EIC should allow one to
measure the $t$ dependence of the DVCS and BH cross sections, which will tremendously
increase the potential to probe nuclear GPDs in the domain of nuclear shadowing (small $x_B$) (see Fig.~\ref{fig:sigma_dvcs_tdep} and the previous discussion).
 
Another possibility to study nuclear GPDs in the small
$x_B$ region is given by the measurement of DVCS cross section asymmetries (with polarized 
lepton beams or with lepton beams with the opposite electric charges), which are 
proportional to the interference between the DVCS and BH amplitudes.
As an example, we consider the DVCS beam-spin asymmetry, $A_{LU}$, 
measured with the polarized lepton beam and an unpolarized target 
(which is always the case for spin-0 nuclei that we consider).
To the leading twist accuracy, the expression for $A_{LU}$ for a spinless
nuclear target reads~\cite{Guzey:2008th,Belitsky:2001ns,Belitsky:2000vk}
\begin{equation}
A_{LU}(\phi)=-\frac{8K (2-y) Z F_A(t) \Im m {\cal H}_{A}(\xi_N,t,Q^2) \sin \phi }{\frac{1}{x_B}
|{\cal A}_{\rm BH}(\xi_N,t,Q^2,\phi)|^2+\frac{x_B \,t \,{\cal P}_1(\phi){\cal P}_2(\phi)}{Q^2} 4(1-y+y^2/2)|\Im m {\cal H}_{A}(\xi_N,t,Q^2)|^2
} \,,
\label{eq:ALU}
\end{equation}
where $K \propto \sqrt{t_{\rm min}-t}$ is the kinematic factor~\cite{Belitsky:2001ns}, 
$Z$ is the nuclear charge, $\Im m {\cal H}_{A}$ is the imaginary part of the nuclear
DVCS amplitude given by Eqs.~(\ref{eq:singlet1}), (\ref{eq:singlet2}) and
(\ref{eq:final}), $|{\cal A}_{\rm BH}(\xi_A,t,Q^2,\phi)|^2$ is the square of the BH
amplitude~(\ref{eq:BH}), and the minus in front corresponds to the electron beam.
To consistently work to the leading twist accuracy, one should use only the
leading twist contributions to ${\cal P}_1(\phi)$, ${\cal P}_2(\phi)$ and 
$|{\cal A}_{\rm BH}|^2$ in Eq.~(\ref{eq:ALU}). However, in the kinematics that we consider, $t < 0.2$ GeV$^2$, $Q^2=2.5$ GeV$^2$ and $\phi=90^{\circ}$, 
the higher twist corrections are either absent (the terms being proportional to
$\cos \phi$) or numerically insignificant, so that
we simply use the standard expressions for 
${\cal P}_1(\phi)$, ${\cal P}_2(\phi)$ and 
$|{\cal A}_{\rm BH}|^2$~\cite{Belitsky:2001ns}.

\begin{figure}[t]
\begin{center} 
\epsfig{file=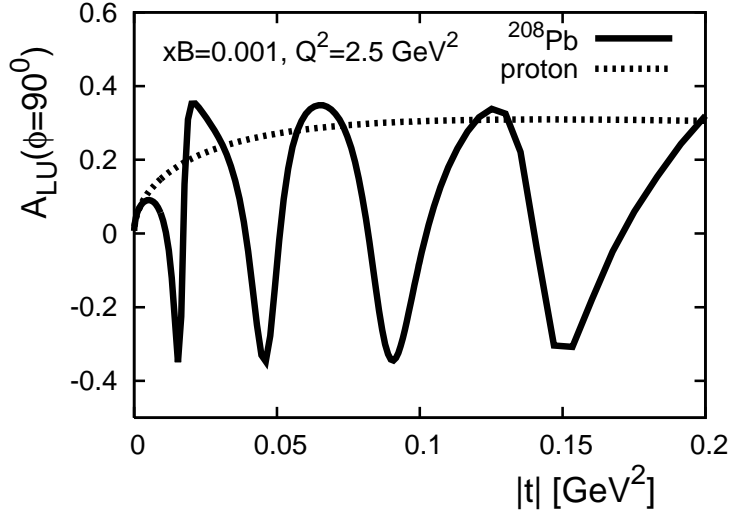,scale=1.3}
\caption{The DVCS beam-spin asymmetry, $A_{LU}(\phi=90^{\circ})$, for $^{208}$Pb
as a function of $t$ at fixed $x_B=0.001$ and  $Q_0^2=2.5$ GeV$^2$
(solid curve). For a comparison, the dotted curve presents
$A_{LU}$ for the proton in the same kinematics. The calculations correspond to
$y=0.31$.
}
\label{fig:ALU_tdep}
\end{center}
\end{figure}

Figure~\ref{fig:ALU_tdep} presents our predictions for $A_{LU}(\phi)$
as a function of $t$ as fixed $x_B=0.001$, $Q_0^2=2.5$ GeV$^2$, and the angle
$\phi=90^{\circ}$. For a comparison, the dotted curve presents
$A_{LU}$ for the proton in the same kinematics. Both curves correspond to
the incoming lepton fractional energy loss $y=0.31$, which in turn corresponds
to the high-energy option of the future EIC with 
$E_{\rm lepton}=20$ GeV and $E_{\rm nucleus}=100$ GeV/nucleon.

Our predictions for $A_{LU}$ for $^{208}$Pb are rather remarkable.
The sole reason for the oscillations of $A_{LU}$ for $^{208}$Pb is nuclear shadowing!
The trend of the oscillations can be understood as follows.
At $t=t_{\min}$, $A_{LU}=0$ because of the kinematic factor $K=0$ (resulting from the 
vanishing $|\vec{\Delta}_{\perp}|=0$).
As one slightly increases $|t|>|t_{\min}|$, the kinematic factors
rapidly increase $A_{LU}$ (which is clearly seen for the proton), but, at the same time,
the nuclear shadowing correction decreases the imaginary part of the nuclear
DVCS amplitude, $\Im m {\cal H}_{A}$. As a result, $A_{LU}$ increases, but not as rapidly as
for the free proton case.
At some rather small values of $t$, $|t| \approx 0.01$
GeV$^2$ (a value that can be read off the left panel of Fig.~\ref{fig:R_xixi_fixed_t}),
$\Im m {\cal H}_{A}$ changes sign and $A_{LU}$ goes through zero.
Note that at this values of $t$, the nuclear form factor, $F_A(t)$, is still positive.
As one increases $|t|$ further, $|\Im m {\cal H}_{A}|$ increases, which increases
$|A_{LU}|$ (with both $\Im m {\cal H}_{A}$ and $A_{LU}$ being negative at this point).
As $|t|$ is increased even further, the nuclear form factor $F_A(t)$ changes sign and
makes $A_{LU}$ positive. The asymmetry stays positive until $\Im m {\cal H}_{A}$ changes sign
and becomes positive again [the form factor $F_A(t)$ still being negative].
As $|t|$ is increased, the mechanism of the oscillations just described repeats itself.

We emphasize that the oscillations of $A_{LU}$ are caused by nuclear shadowing
that has a weaker $t$ dependence than that of the Born contribution
[see Eq.~(\ref{eq:final})]. If the shadowing correction in Eq.~(\ref{eq:final}) is neglected,
then the $t$ dependence of the DVCS and BH contributions is the same and is given by the
nuclear form factor $F_A(t)$. Then, in the expression for the beam spin asymmetry, 
$A_{LU}$, the $t$ dependence from $F_A(t)$ cancels and $A_{LU}$ for a heavy nuclear
target has the same $t$ dependence as $A_{LU}$ for the free proton
(i.e., without the oscillations).

\section{Summary and discussion}
\label{sec:summary}

We generalized the leading twist theory of nuclear shadowing for usual nuclear parton
distributions to nuclear generalized parton distributions  for quarks and gluons.
We estimated quark and gluon GPDs of spinless nuclei and found very large nuclear
shadowing. 

In the limit that the momentum transfer is purely transverse, $\xi_A=\xi_N=0$,
after Fourier transform, our nuclear GPDs become impact-parameter-dependent
nuclear PDFs. Nuclear shadowing induces non-trivial correlations between the
impact parameter $b$ and the light-cone fraction $x$.

Using our expressions for nuclear GPDs, we made predictions for the
cross section of deeply virtual Compton scattering
 on the heavy nucleus of $^{208}$Pb at high energies (in the
kinematics of the future EIC).
We also calculated the cross section of the purely electromagnetic 
Bethe-Heitler process and addressed the issue of the extraction of the
DVCS signal, and, hence, the extraction of information on nuclear GPDs and
nuclear shadowing, from the measurement of the $e A \to e \gamma A$ process.
Based on our studies, we can propose two strategies. First, 
the $e A \to e \gamma A$ differential cross section at the momentum
transfer $t$ near the minima of the nuclear form factor is dominated by the DVCS
cross section, which should allow for a clear extraction 
of the latter. Second, nuclear shadowing leads to dramatic oscillations
of the DVCS beam-spin asymmetry, $A_{LU}$, as a function of $t$.
The position of the points where $A_{LU}$
changes sign is directly related to the magnitude of nuclear shadowing.

It is important to note that the $t$ variations of the DVCS and BH differential cross
sections and the DVCS beam-spin asymmetry, $A_{LU}$, are very rapid, with the typical
frequency of the order of $1/R_A^2$. 
This certainly poses a challenge for the future experiment since
a rather high resolution in $t$ will be required.

One should also note that nuclear GPDs at small $x$ will be accessed 
in ultraperipheral nucleus-nucleus collisions at the LHC~\cite{Baltz:2007kq}.
 In these collisions,
the involved nuclei serve as sources of real photons, which
enables one to study photon-nuclear interactions at energies up to ten times larger than 
those achieved at HERA. Nuclear GPDs will be accessed in exclusive photoproduction of
heavy vector mesons~\cite{Frankfurt:2003qy} and lepton pairs~\cite{Pire:2008ea}.

\acknowledgments

We would like to thank M.~Strikman for useful discussions.
This paper is
authored by Jefferson Science Associates, LLC under U.S. DOE Contract No. DE-AC05-06OR23177. The U.S. Government retains a non-exclusive, paid-up, irrevocable, world-wide license to publish or reproduce this manuscript for U.S. Government purposes.

\end{document}